\documentclass[10pt,journal]{IEEEtran}
\usepackage{amsmath,amssymb,amsfonts}
\usepackage{graphicx}
\usepackage{multirow}
\usepackage{cuted}
\usepackage{CJK}
\usepackage{indentfirst}
\usepackage{xcolor}
\usepackage{hyperref}
\usepackage{epstopdf}
\usepackage{bm}
\usepackage{cases}
\usepackage{subfigure}
\usepackage{caption}
\usepackage{bookmark}
\usepackage{makecell}

\usepackage{algorithmic}
\usepackage{algorithm}
\usepackage{array}
\usepackage{textcomp}
\usepackage{stfloats}
\usepackage{url}
\usepackage{verbatim}
\usepackage{cite}
\hypersetup{
colorlinks=true,
linkcolor=black,
citecolor=black
}
\begin{document}
\title{Joint beamforming and compressed sensing for uplink grant-free access}
\author{Guoqing Xia, Pei Xiao, {\it Senior Member, IEEE}, Bohan Li, Yue Zhang, {\it Senior Member, IEEE}, Huiyu Zhou
\thanks{This work was supported in part by EU Horizon 2020 project 6G BRAINS under Grant
101017226 (Corresponding author: Yue Zhang).}
\thanks{Guoqing Xia is with the School of Engineering, University of Leicester, LE1 7RH
Leicester, UK (e-mail: gx21@leicester.ac.uk).}
\thanks{Pei Xiao is with 5GIC $\&$ 6GIC, Institute for Communication Systems (ICS) of University of Surrey,
Guildford, GU2 7XH, UK (e-mail: p.xiao@surrey.ac.uk).}
\thanks{Bohan Li and Huiyu Zhou are with the School of Computing and Mathematical Sciences, University of Leicester, LE1 7RH
Leicester, UK (e-mail: bl204@leicester.ac.uk and hz143@leicester.ac.uk).}
\thanks{Yue Zhang is with the Institute of Communication Measurement Technology, Chengdu 610095, China (e-mail: zhangyue@icsmcn.cn).}}
\markboth{IEEE TRANSACTIONS ON WIRELESS COMMUNICATIONS, Vol. 23, No. 09, SEPTEMBER 2024}
{Shell \MakeLowercase{\textit{et al.}}: Bare Demo of IEEEtran.cls for IEEE Journals}
\maketitle
\begin{abstract}
Compressed sensing (CS)-based techniques have been widely applied in the grant-free non-orthogonal multiple access (NOMA) to a single-antenna base station (BS). In this paper, we consider the multi-antenna reception at the BS for uplink grant-free access for the massive machine type communication (mMTC) with limited channel resources. To enhance the overloading performance of the BS, we develop a general framework for the synergistic amalgamation of the spatial division multiple access (SDMA) technique with the CS-based grant-free NOMA. We derive a closed-form statistical beamforming and a dynamic beamforming scheme for the inter-cluster interference suppression when applying SDMA. Based on this, we further develop a joint adaptive beamforming and subspace pursuit (J-ABF-SP) algorithm for the multiuser detection and data recovery, with a novel sparsity level decision method without the accurate knowledge of the noise level. To further improve the data recovery performance, we propose an interference cancellation-based J-ABF-SP scheme (J-ABF-SP-IC) by using the initial signal estimates generated from the J-ABF-SP algorithm. Illustrative simulations verify the superior user detection and signal recovery performance of our proposed algorithms in comparison with existing CS-based grant-free NOMA techniques.
\end{abstract}
\begin{IEEEkeywords}
   mMTC, Grant-free access, NOMA, Beamforming, Subspace pursuit, Joint optimisation, Interference cancellation.
\end{IEEEkeywords}

\section{Introduction}\label{sec:Intro}
The {\it massive machine type communication} (mMTC), e.g., the internet of things (IoT), emerged in the 5G era, will still play a critical role in the forthcoming beyond 5G and even 6G eras. {\it Non-orthogonal multiple access} (NOMA) has been identified as an enabler to support the massive connectivity with limited channel resources \cite{Hoshyar2008, Nikopour2013,Zhang2014,Islam2017,Kusaladharma2021}. Another characteristic of mMTC is sporadic data transmission, i.e., at any time only a small fraction of potential users are active and transmit small data packets \cite{Shariatmadari2015,3GPP_GF,Dawy2017,Liu2018MSP}. In this case, the conventional grant-based NOMA techniques will cause the large access delay and signalling overhead. Therefore, an efficient communication paradigm shift is necessary to enable the low-latency and high-reliability mMTC applications.
\subsection{Related Work}\label{sec: related}
Recently, {\it grant-free NOMA} methods have been envisioned as feasible solutions for mMTC. In the uplink grant-free access, the active users transmit data via the available channel resources that the BS broadcasts periodically, without going through the complicated channel access request and granting process \cite{Liu2018MSP,Shahab2020}. Thus, the grant-free access is effective in reducing the access delay and signalling overhead due to the sporadic and small-scale data transmission in the mMTC scenario. However, in the grant-free access, the BS cannot identify the active users before data transmission without the granting process. Thus, for reliable uplink communications, blind user activity detection is necessary via the superimposed received signal of the active users.

\begin{figure}[!t]
   \centerline
   {\includegraphics[width=0.2\textwidth]{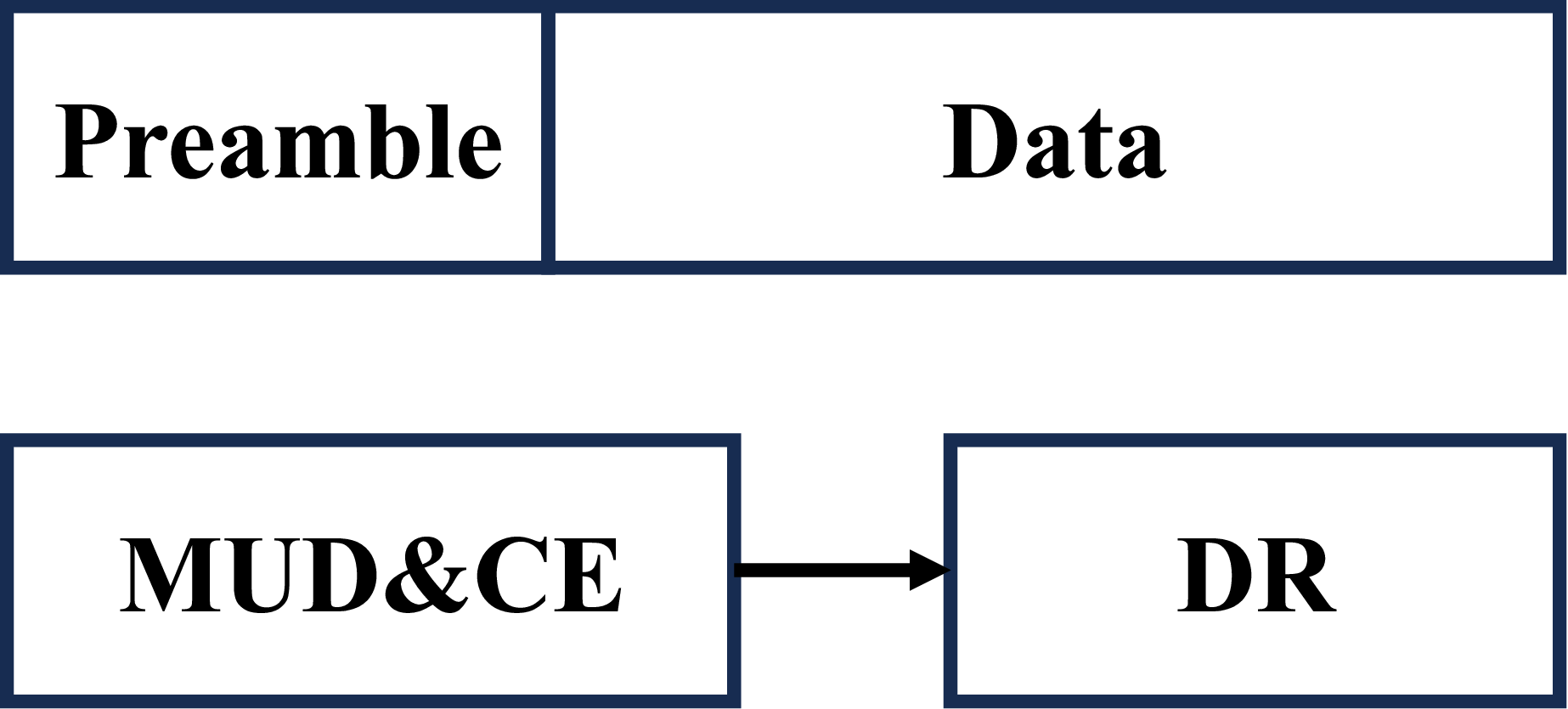}}
   \caption{Frame structure of the first grant-free access type}\label{fig:Frame_GF1}
\end{figure}

\begin{figure}[!t]
   \centerline
   {\includegraphics[width=0.2\textwidth]{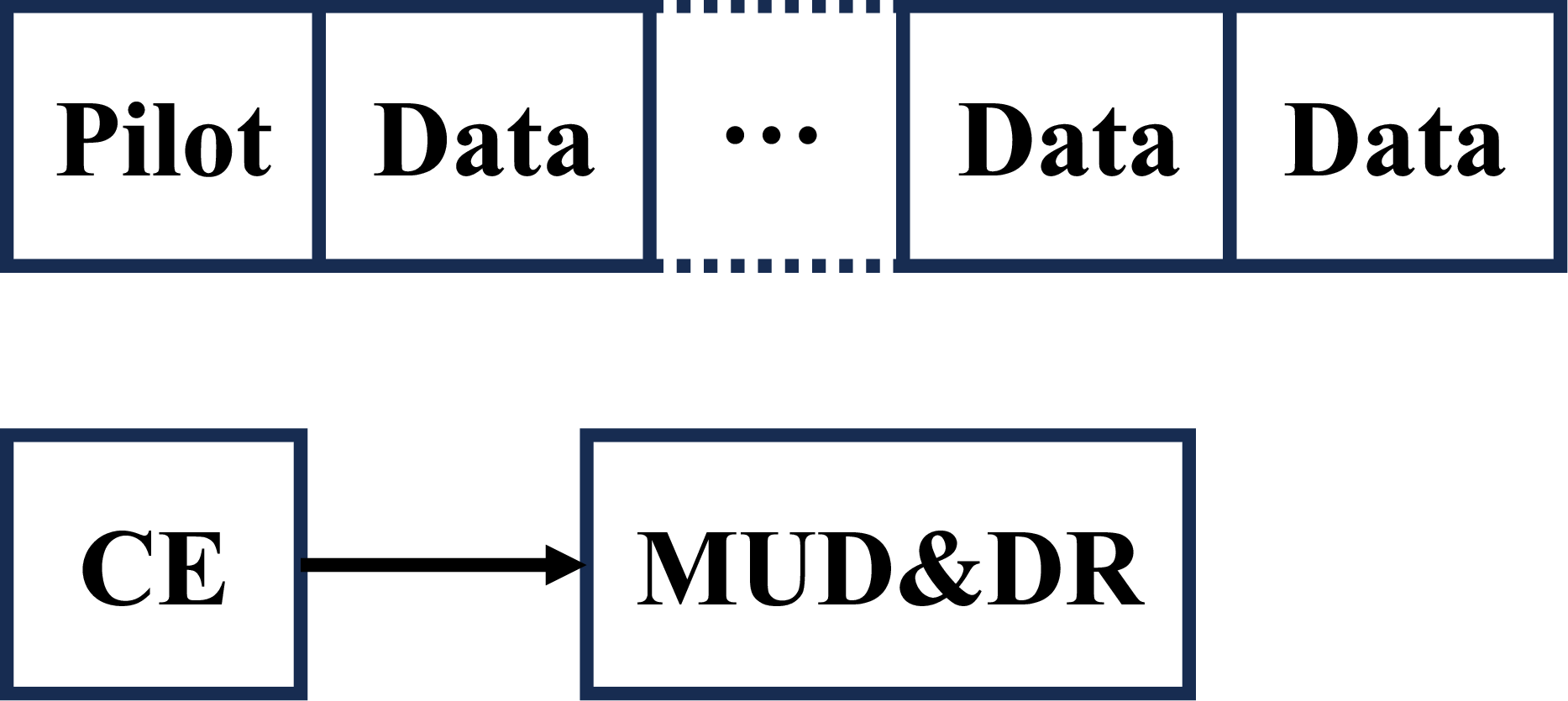}}
   \caption{Frame structure of the second grant-free access type}\label{fig:Frame_GF2}
   \vspace{-0.3cm}
\end{figure}

Current coherent grant-free access schemes can be classified into two categories according to the method of channel estimation and user activity detection \cite{Qiao2022}. For the first grant-free access type, the preambles of the active users are transmitted to the BS for channel estimation (CE) and multiple user detection (MUD), and the coherent data recovery (DR) is then performed at the BS based on the previously estimated channel state information \cite{Senel2018,Liu2018,Ahn2019,Ke2020}. For the second grant-free access type, the channel information of all the users are estimated based on pilots in the first stage, and subsequently within the coherence time, the joint MUD and DR is performed at the BS \cite{Wei2017AMP,Du2018,Gao2022,Mei2022}. The frame structures of these two grant-free schemes are shown in Figs. \ref{fig:Frame_GF1} and \ref{fig:Frame_GF2}.
In addition, some non-coherent grant-free access methods are proposed for some specific applications, e.g., unmanned aerial vehicle (UAV) assisted massive IoT \cite{Qiao2022} and massive multiple-input-multiple-output (MIMO) \cite{Senel2018}. In this paper, we focus on the joint MUD and DR for the second type of grant-free access for mMTC.

The sporadic transmission in mMTC gives rise to the sparse received signal with high probability. {\it Compressed sensing} (CS) techniques are promising in recovering the sparse signals from the far fewer samples than those required by the classic Nyquist sampling \cite{Tropp2007,Guo2008,Needell2009,Dai2009,donoho2009message}. Accordingly, the number of necessary resource elements for data transmission can be reduced when considering the CS-based receiver. The CS-based grant-free NOMA necessitates judicious transceiver design. At the transmitter, the active users modulate the information bits into symbols, and spread them onto specific subcarriers by using non-orthogonal signatures for transmissions. The widely used spreading schemes include low density signature (LDS) \cite{Hoshyar2008}, sparse code multiple access (SCMA) \cite{Nikopour2013,Zhang2014,Luo2022,Luo2022Non_coherent}, etc.. At the receiver, the received signals on different subcarriers are used for the user activity detection and signal recovery by CS techniques. Extensive CS-based sparse signal recovery methods have been proposed, including the orthogonal matching pursuit (OMP) \cite{Tropp2007}, compressed sampling matching pursuit (CoSaMP) \cite{Needell2009}, subspace pursuit (SP) \cite{Dai2009} and approximate message passing (AMP) method \cite{donoho2009message}, etc.. These methods require prior knowledge of the user sparsity level (the number of active users), which is often impractical in engineering applications. 

Furthermore, considering the consecutive data transmission in different slots in mMTC scenarios, the temporal correlation for the user activity has been utilised to enhance the communication performance in grant-free NOMA systems \cite{Wei2017AMP,Du2018,Gao2022,Mei2022, Wang2016,Du2017,Li2022,Wu2022}. The assumptions on the temporal correlation of the user activity can be classified into two categories. The first one is that the user activity stays unchanged in one frame, called {\it frame-wise (block) sparsity}. Based on this assumption, the modified AMP \cite{Wei2017AMP}, SP \cite{Du2018} and block-coordinate-descent (BCD) \cite{Gao2022} methods were developed for the frame-wise user activity detection and data recovery in grant-free NOMA. These methods do not require the prior user sparsity level but need to estimate it based on the prior noise power. To avoid using the prior information of the noise level, the authors in \cite{Du2018} proposed a cross-validation-based method to determine the user sparsity level. The authors in \cite{Mei2022} considered an orthogonal approximate message passing (OAMP)-multiple measurement vector (MMV) algorithm with simplified structure learning (SSL) and accurate structure learning (ASL), termed as OAMP-MMV-SSL and OAMP-MMV-ASL, respectively. These two methods can iteratively estimate the user sparsity ratio and the noise variance using the expectation maximisation \cite{Mei2022}. 

The second is the dynamic user sparsity assumption, i.e., the user activity can be different in consecutive slots. A dynamic CS method \cite{Wang2016} and a modified SP method \cite{Du2017} were proposed to improve the active user estimates in consecutive slots based on the temporal correlation between one another. The weighted $l_{2,1}$ minimisation model-based method was developed for the enhanced performance in detecting the users with dynamic sparsity \cite{Li2022}. In addition, the first bit with value 0 or 1 in the data payload was used to determine whether the active user has data to transmit in the current time slot \cite{Wu2022}. All of these methods require the noise level as the prior information. 

The aforementioned methods are usually developed for the grant-free NOMA system with a {\it single-antenna BS}. Recently, \cite{Liu2018} demonstrated that, both the missed user detection and the false alarm probabilities can always converge to zero by utilising the vector AMP algorithm \cite{donoho2009message}, in the asymptotic massive MIMO regime. A joint spatial-temporal-structured adaptive SP method was proposed for grant-free NOMA to jointly estimate channels and detect users by considering the block sparsity over multiple slots and multiple antennas \cite{Wu2021}. Additionally, media-based modulation is employed in grant-free access in multi-antenna BS scenarios by using SP \cite{Ma2019,Qiao2020} and AMP \cite{Qiao2022MM}. However, these spatial modulation methods do not fully exploit the inherent spatial diversity and multiplexing gain of the potential user clustering and thus require a large number of antennas to achieve a satisfactory performance.
\subsection{Motivation}
Accurate sparse signal recovery necessitates a large number of spectrum resources or massive antennas for massive connectivity with current CS-based grant-free NOMA techniques, even though they can enable the system to operate in overloaded conditions to some extent \cite{Wei2017AMP,Du2018,Gao2022,Mei2022,Ma2020,Qiao2020,Qiao2022MM}. The {\it spatial division multiple access} (SDMA) technique characterised by the {\it multi-antenna BS} has been proven to be effective in supporting massive connectivity, especially when integrating with the power-domain NOMA techniques \cite{Xu2017,Cui2018, Zhu2019,Senel2019,Zhang2020,Xia2022}. As shown in Fig. \ref{fig:SDMA}, the SDMA can cope with the simultaneous transmissions of multiple users sharing the same spectrum resources aided by an advanced interference mitigation technique, e.g., digital beamforming. It is a promising solution to integrate the SDMA with the CS-based grant-free NOMA technique in mMTC applications for improved spectral efficiency. However, to our best knowledge, there is no work in the open literature that has taken this into consideration.

\subsection{Our Contribution}
In this paper, we concentrate on developing the joint MUD and DR method for the uplink grant-free NOMA to a multi-antenna BS. We consider i) the first temporal correlation assumption, i.e., the frame-wise block sparsity for each user; ii) the second coherent grant-free access type with the channel information estimated using pilots before the data transmission. Massive users are assumed to be clustered according to the channel correlation, based on which the multi-antenna reception can be combined by beamforming to suppress the inter-cluster interferences. For users within the same cluster, the CS-based grant-free NOMA method is utilised for the MUD and DR based on the combined signal by beamforming.  The main contributions are summarised as follows. 

1) We have developed both a closed-form statistical beamforming (SBF) scheme and a dynamic beamforming (DBF) scheme. These beamforming approaches, when combined with appropriate user clustering based on channel correlation, effectively mitigate inter-cluster interferences. Even in cases where the total number of users significantly exceeds the number of antenna elements at the base station, these schemes demonstrate effective interference suppression.

2) We have formulated a comprehensive framework for integrating SDMA with grant-free NOMA. This framework enables simultaneous differentiation and service of spatially clustered users using the spatial diversity and multiplexing gain provided by multiple beams. Within this structure, the optimisation of beamforming and signal estimation is jointly and alternately performed. This parallel optimisation process for distinct user clusters can significantly reduce the access latency. Additionally, the utilisation of the same spectrum resources by all user clusters leads to a substantial increase in spectral efficiency.

3) As a practical realisation of the developed framework, we introduce a joint adaptive beamforming and subspace pursuit (J-ABF-SP) algorithm tailored for uplink grant-free access. In each iteration of the J-ABF-SP algorithm, adaptive beamforming and subspace pursuit are performed alternately to jointly achieve user detection and signal recovery. A robust method for determining user sparsity level is introduced, obviating the need for prior knowledge of noise levels.

4) To further enhance MUD and DR performance, we propose an interference cancellation (IC) scheme denoted as J-ABF-SP-IC. Building upon the results obtained from user activity detection and initial signal estimation via the J-ABF-SP algorithm, this scheme involves the reconstruction of received signals for each cluster. By utilising these reconstructed signals, interference-cancelled received signals for each cluster are derived. Subsequently, similar procedures to those in the J-ABF-SP algorithm are used to alternate between signal estimation and beamforming optimisation.

5) Simulation results verify that the J-ABF-SP algorithm can achieve superior MUD and DR performance in comparison with the benchmark methods at the cost of moderately increased complexity. Moreover, the J-ABF-SP-IC algorithm can further enhance the performance with slightly increased complexity. In addition, compared to the existing methods, the integration of the SDMA and grant-free NOMA in this paper can markedly improve the spectral efficiency.

The remainder of the following parts of this paper is organised as follows. Section II describes the signal model and
problem formulation. Section III introduces the proposed beamforming schemes. Section IV details the proposed joint optimisation algorithms for the beamforming and data recovery.
Section V gives the computational complexity analysis. Section VI illustrates the simulation results. Section VII concludes this paper.

\emph{Notation}: $\mathbb{C}$ denotes the field of complex numbers. Scalars are denoted by lower-case letters, vectors and matrices respectively by lower- and upper-case boldface letters. The conjugate, transpose, conjugate transpose and Moore-Penrose (M-P) inverse are denoted by $(\cdot)^*$, $(\cdot)^{\rm T}$ $(\cdot)^{\rm H}$ and $(\cdot)^{\dagger}$, respectively. $\mathbb{E}\{\cdot\}$ and $|\cdot|$ denote the mathematical expectation and modulus, respectively. ${\rm vec}\{\cdot\}$ vectorises a matrix by stacking each column of it on top of one another. ${\rm vec}^{-1}({\bm c},\mathcal{T})$ generates a matrix with $\mathcal{T}$ rows by performing inversely vectorisation to the vector ${\bm c}$. $\|\cdot\|_{2}$ denotes the $l_2$ norm of a matrix. $\|\cdot\|_{0}$ denotes the $l_0$ norm of a vector, i.e., the number of non-zero elements of it. The notations $\min\{\cdot\}$ and $\max\{\cdot\}$ denote the minimum and maximum element of the enclosed set $\{\cdot\}$, respectively. The notation $\otimes$ denotes the Kronecker product.

\section{Signal Model and Problem Formulation}\label{Sec:syst model}
\begin{figure}[!t]
   \centerline
   {\includegraphics[width=0.3\textwidth]{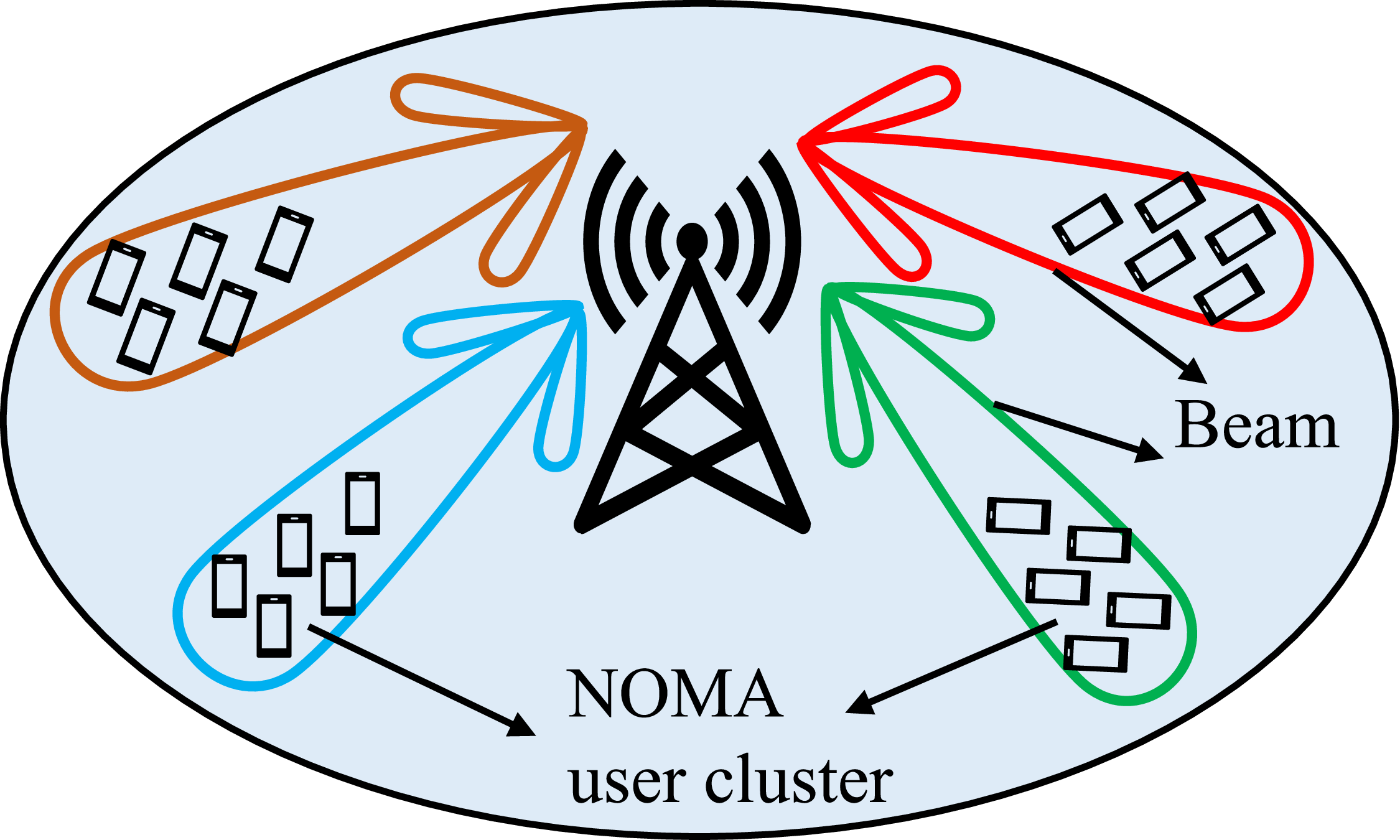}}
   \caption{System architecture of the integration of SDMA and grant-free NOMA}\label{fig:SDMA}
   \vspace{-0.3cm}
\end{figure}

We consider the spreading-based grant-free NOMA in a multi-antenna cellular system to support the mMTC with limited channel resources. The cellular BS is equipped with a uniform linear array with $M$ antenna elements while all users are with a single antenna. We consider the second coherent grant-free access type with the channel information estimated using pilots before the data transmission, as illustrated in Fig. \ref{fig:Frame_GF2}. As shown in Fig. \ref{fig:SDMA}, $NQ$ users (devices) are grouped into $N$ clusters \footnote{The use cases involve Industry IoT, e.g., a smart factory  where lots of sensors perform some monitoring and transmission tasks and sensors in the close direction can be clustered for grant-free access.} according to their channel correlation by using common clustering methods, e.g., K-means~\cite{Cui2018,Zhang2020,Le2021}. The channel correlation coefficient is defined in Appendix \ref{sec:channel correlation}. Without loss of generality, the equal-size clusters are assumed, e.g., $Q$ users in each cluster $n=1,2,\cdots,N$.  All user clusters employ the same frequency resources, i.e., $K$ subcarriers, for simultaneous communication with the BS. To support mMTC, we consider an overloaded system with $K<NQ$\footnote{In fact, $K<Q$ can be satisfied since we consider all clusters use the same spectrum resource.}. 

Please note that the number of user clusters is constrained by the degrees-of-freedom (DoF) of the BS, while the angular distribution range of users in each cluster is limited by the main lobe width of the beampattern. Both the DoF and the main lobe width of the beampattern are determined by the number of antenna elements in a specific array configuration. Consequently, for a given user distribution, the number of user clusters and the angular distribution range of users in each cluster should match the number of antennas. This ensures sufficient utilisation of the spatial resources and helps prevent the performance degradation.

To enhance the readability of the signal model and algorithm derivations, we provide a summary of the key variables involved in Table \ref{tab:variables}. This table includes their definitions and dimensions for clarity.
\begin{table*}[!t]
       \centering
       \footnotesize
       \caption{A summary of the key variables in this paper}
       \vspace{-0.1cm}
      \label{tab:variables}	
       \begin{tabular}{|l|c|}
           \hline
           Variable&Definition\\\hline
           $x_{n,q,t}\in\mathbb{C}$&the transmitted signal of user $u_{n,q}$ at the current slot $t$\\\hline
           ${\bm x}_{n,t}\in\mathbb{C}^{Q\times 1}$&the transmitted signal vector with its $q$th entry being $x_{n,q,t}$\\\hline
           ${\bm X}_{n}\in\mathbb{C}^{Q\times \mathcal{T}}$&the transmitted signal matrix for cluster $n$, its $q$th column being ${\bm x}_{n,t}$\\\hline
           $\tilde{\bm G}_{n,k}\in\mathbb{C}^{M\times Q}$&the equivalent channel gain matrix for cluster $n$ on subcarrier $k$\\\hline
           ${\bm y}_{k,t}\in\mathbb{C}^{M\times 1}$&the received signal at the BS on subcarrier $k$ and at slot $t$, formulated in \eqref{eq:received_k}\\\hline
           ${\bm b}_{n}\in\mathbb{C}^{M\times 1}$&the beamforming weight vector for cluster $n$\\\hline
           $y_{n,k,t}\in\mathbb{C}$&the combined received signal for cluster $n$ on subcarrier $k$, formulated in \eqref{eq:BF_received_k}\\\hline
           ${\bm y}_{n,t}\in\mathbb{C}^{K\times 1}$&the combined received signal vector for cluster $n$, with its $k$th entry being $y_{n,k,t}$\\\hline
           ${\bm y}_t\in\mathbb{C}^{MK\times 1}$&the received signal vector formed by cascading ${\bm y}_{k,t}$ for all $k$ in column\\\hline
           $\tilde{\bm G}_{n}\in\mathbb{C}^{MK\times Q}$&the equivalent channel matrix by cascading $\tilde{\bm G}_{n,k}$ for all $k$ in column\\\hline
           ${\bm B}_{n,l}\in\mathbb{C}^{K\times Q}$&the equivalent beamforming gain matrix, formulated in \eqref{eq:B_nn}\\\hline
           ${\bm Y}\in\mathbb{C}^{MK\times \mathcal{T}}$&the received signal matrix formed by cascading ${\bm y}_{t}$ for all $t$ in row, formulated in \eqref{eq:received_matrix}\\\hline
           ${\bm Y}_n\in\mathbb{C}^{K\times \mathcal{T}}$&the combined received signal matrix for cluster $n$ formed by cascading ${\bm y}_{n,t}$ for all $t$ in row, formulated in \eqref{eq:BF_received_mat}\\\hline
           ${\bm \eta}_n\in\mathbb{C}^{K\mathcal{T}\times 1}$, ${\bm c}_{n}\in\mathbb{C}^{Q\mathcal{T}\times 1}$&the vectorisations of ${\bm Y}_n$ and ${\bm X}_n$, respectively\\\hline
           ${\bm {\mathcal{D}}}_{n}\in\mathbb{C}^{K\mathcal{T}\times Q\mathcal{T}}$&the parameter matrix for cluster $n$ formed by ${\bm B}_{n,n}$, defined in \eqref{eq:received vectorisation}\\\hline
      \end{tabular}
   \vspace{-0.3cm}
\end{table*}
\subsection{Signal Model}\label{Sec:sig model}

The $q$th user in cluster $n$ is expressed by $u_{n,q}$. The spreading signature for $u_{n,q}$ is denoted as ${\bm s}_{n,q}=[s_{n,q,1},s_{n,q,2},\cdots,s_{n,q,K}]^{\rm T}$ with $s_{n,q,k}$ representing the spreading factor on subcarrier $k$ for user $u_{n,q}$ \cite{Gao2022,Mei2022,Li2022}. Non-orthogonal non-sparse spreading signatures are employed in this paper, e.g., Zadoff-Chu sequences \footnote{The Zadoff-Chu spreading signatures are detailed in Appendix \ref{sec:Appen_Zadoff}.} \cite{Popovic1992}. Assuming the line-of-sight transmission only, the angle of arrival (AoA) from user $u_{n,q}$ can be denoted as $\theta_{n,q}$ and the steering vector is defined as,
\begin{align}
   {\bm a}_{n,q} = \begin{bmatrix}1& e^{-j2\pi\dfrac{d\sin(\theta_{n,q})}{\lambda}}& \cdots & e^{-j2\pi(M-1)\dfrac{d\sin(\theta_{n,q})}{\lambda}}\end{bmatrix}^{\rm T}\label{eq:steering vector}
\end{align}  
where $e$ is the Euler's number, $\lambda$ is the carrier wavelength and $d$ is the distance between the adjacent antenna elements, usually set to be a half wavelength $\lambda/2$.
The channel gain vector ${\bm g}_{n,q,k}\in\mathbb{C}^{M\times 1}$ between the user $u_{n,q}$ and the multi-antenna BS using subcarrier $k$ can be modelled as the product of the channel fading and the steering vector, defined as ${\bm g}_{n,q,k} = f_{n,q,k}{\bm a}_{n,q}$, where the channel fading $f_{n,q,k}=\rho_{n,q}\eta_{n,q,k}$ consists of the large-scale fading $\rho_{n,q}$, including the path loss and shadowing fading, and the small-scale random fading $\eta_{n,q,k}$ following the complex Gaussian distribution. We assume a slow-fading channel which remains unchanged within a coherence time interval (longer than the frame length of the mMTC). 

The received signal at the BS for subcarrier $k$ and slot $t$ can be formulated as,
\begin{align}
   {\bm y}_{k,t} &= \sum\nolimits_{n=1}^N\sum\nolimits_{q=1}^Q{\bm g}_{n,q,k}s_{n,q,k}x_{n,q,t}+{\bm v}_{k,t}\notag\\
   &=\sum\nolimits_{n=1}^N\tilde{\bm G}_{n,k}{\bm x}_{n,t}+{\bm v}_{k,t},\label{eq:received_k}
\end{align}
where $x_{n,q,t}$ \footnote{Note that $x_{n,q,t}=0$ for each inactive user $u_{n,q}$.} is the transmitted signal of user $u_{n,q}$ at the current slot $t$, ${\bm x}_{n,t}$ is the transmitted signal vector with its $q$th entry being $x_{n,q,t}$, and ${\bm v}_{k,t}$ is the additive Gaussian noise vector. The equivalent channel gain matrix for cluster $n$ on subcarrier $k$ is $\tilde{\bm G}_{n,k}\triangleq [\tilde{\bm g}_{n,1,k},\tilde{\bm g}_{n,2,k},\cdots,\tilde{\bm g}_{n,Q,k}]\in\mathbb{C}^{M\times Q}$ with the equivalent channel gain vector $\tilde{\bm g}_{n,q,k}\triangleq s_{n,q,k}{\bm g}_{n,q,k},\ q=1,2,\cdots,Q$.

Since the users are clustered by channel correlation, beamforming can be performed to suppress the inter-cluster interference signals at the BS. For any cluster $n=1,2,\cdots,N$, the multi-antenna received signal on subcarrier $k$ is combined by beamforming, i.e.,
\begin{align}
   y_{n,k,t} &={\bm b}_{n}^{\rm H}{\bm y}_{k,t}\label{eq:BF_received_k} 
   =\sum\nolimits_{l\in\mathcal{N}}{\bm b}_{n}^{\rm H}\tilde{\bm G}_{l,k}{\bm x}_{l,t}+{\bm b}_{n}^{\rm H}{\bm v}_{k,t},
\end{align}
where $\mathcal{N}$ is the index set of all clusters, and ${\bm b}_{n}$ is the beamforming weight vector for cluster $n$.

Cascading $y_{n,k,t}$ by $k=1,2,\cdots,K$ yields the combined signal vector ${\bm y}_{n,t}\in\mathbb{C}^{K\times 1}$, 
\begin{align}
   {\bm y}_{n,t} &=\left({\bm I}_K\otimes {\bm b}_{n}\right)^{\rm H}{\bm y}_t\label{eq:BF_received}
\end{align}
where ${\bm I}_K$ denotes a $K\times K$ identity matrix and the received signal vector ${\bm y}_t$ is given by, 
\begin{align}
   {\bm y}_t&=\left[{\bm y}^{{\rm T}}_{1,t},{\bm y}^{{\rm T}}_{2,t},\cdots,{\bm y}^{{\rm T}}_{K,t}\right]^{\rm T}= \sum\nolimits_{n=1}^N\tilde{\bm G}_{n}{\bm x}_{n,t}+{\bm v}_{t}, \label{eq:received_vector}
\end{align}
with the equivalent channel matrix $\tilde{\bm G}_{n}\triangleq \left[\tilde{\bm G}_{n,1}^{{\rm T}},\tilde{\bm G}_{n,2}^{{\rm T}},\cdots,\tilde{\bm G}_{n,K}^{{\rm T}}\right]^{\rm T}\in\mathbb{C}^{KM\times Q}$ and the noise vector ${\bm v}_{t}\triangleq \left[{\bm v}_{1,t}^{{\rm T}},{\bm v}_{2,t}^{{\rm T}},\cdots,{\bm v}_{K,t}^{{\rm T}}\right]^{\rm T}$. 
We define the equivalent beamforming gain matrix,
\begin{equation}
{\bm B}_{n,l}\triangleq \left({\bm I}_K\otimes {\bm b}_{n}\right)^{\rm H}\tilde{\bm G}_{l}\in\mathbb{C}^{K\times Q}. \label{eq:B_nn}
\end{equation}
 Then, ${\bm y}_{n,t}$ can be rewritten as,
\begin{equation}
{\bm y}_{n,t}={\bm B}_{n,n}{\bm x}_{n,t}+\sum\nolimits_{l\in\mathcal{N}\setminus n}{\bm B}_{n,l}{\bm x}_{l,t}+\left({\bm I}_K\otimes {\bm b}_{n}\right)^{\rm H}{\bm v}_t.\label{eq:BF_received_unfold}
\end{equation}
The first term on the right-hand side of \eqref{eq:BF_received_unfold} is the desired signal for cluster $n$, the second is the superimposed inter-cluster interference, and the last is the noise term.
\subsection{Problem Formulation}\label{sec:problem}
As stated in Section \ref{sec: related}, we consider the second grant-free access type, i.e., the channel gains are a priori estimated in the first stage \cite{Wei2017AMP,Du2018,Gao2022,Mei2022}. In this context, we consider non-sparse spreading signatures, such as Zadoff-Chu sequences. With the channel information and spreading signatures, one can obtain the equivalent channel gain matrix $\tilde{\bm G}_{l}$. Our objective is to develop an algorithm that optimises both the beamforming weights and the signal estimates concurrently at the BS.

Define the transmitted signal matrix for cluster $n$ as ${\bm X}_{n}\triangleq [{\bm x}_{n,1},{\bm x}_{n,2},\cdots,{\bm x}_{n,\mathcal{T}}]$, with $\mathcal{T}$ denoting the number of slots in one frame.
According to \eqref{eq:BF_received_unfold}, the least-squares (LS) error function for MUD and DR is given by,
\begin{align}
   \mathcal{E}_{\rm LS}\left({\bm b}_{n},{\bm X}_{n}\right)&=\sum\nolimits_{t=1}^\mathcal{T}\|{\bm y}_{n,t}-{\bm B}_{n,n}{\bm x}_{n,t}\|_2^2,\label{eq:cost_LS}
\end{align}
where $(\cdot)_t$ denotes the random realisation at time slot $t$, e.g., ${\bm y}_{n,t}$, ${\bm y}_{k,t}$ and ${\bm x}_{n,t}$.  

To optimise the signal estimation, we need to constrain the beamforming main lobe towards the desired user cluster by the constraint ${\bm b}_{n}^{\rm H}\bar{\bm a}_n=1$ where $\bar{\bm a}_n\triangleq {1}/{Q}\sum\nolimits_{q=1}^Q{\bm a}_{n,q}$ is the average of the steering vectors of the users in cluster $n$. Herein we use the steering vectors rather than the original channel gain vectors to alleviate the impacts of the random channel fading. The joint optimisation problem can be formulated as,
\begin{align}
   {\rm arg}\min\limits_{{\bm b}_{n},{\bm X}_{n}}&\mathcal{E}_{\rm LS} \left({\bm b}_{n},{\bm X}_{n}\right),\label{eq:problem}\\
      {\rm s.t.}\ & \tilde{\Gamma}_{n,1}=\tilde{\Gamma}_{n,2}=\cdots=\tilde{\Gamma}_{n,\mathcal{T} }=\tilde{\Gamma}_n,\notag\\
      \ &|\tilde{\Gamma}_n|\leq \bar{s},\notag\\
      \ & {\bm b}_{n}^{\rm H}\bar{\bm a}_n=1,\notag
\end{align}
where $\tilde{\Gamma}_{n,t}$ denotes the support set of user cluster $n$ at time slot $t$ and $\bar{s}$ is the maximum user sparsity level. For a slow-fading channel, $\bar{\bm a}_n$ can be obtained by $\bar{\bm a}_n= {1}/{Q}\sum\nolimits_{q=1}^Q{\bm g}_{n,q,k}/{\bm g}_{n,q,k}(1)$ for any $k$.
\section{Beamforming Schemes}
The problem in \eqref{eq:problem} belongs to the multivariate high-order nonlinear constrained optimisation problem, which is generally non-polynomial hard (NP-hard) to solve. In this paper, we consider the joint alternating optimisation for the beamforming weight and the signal estimate. To this end, we first design the effective beamforming schemes for inter-cluster interference suppression.
\subsection{Statistical Beamforming Scheme}\label{sec: SBF}
Ideally, the LS error in \eqref{eq:cost_LS} can be converted into the mean squared error (MSE) when three conditions satisfy, i.e., 1) the number of slots (samples) is large enough, 2) the transmitted signals follow stationary distributions and 3) the channel states stay unchanged within a frame. Based on this, we substitute ${\bm y}_{n,t}$ in \eqref{eq:BF_received_unfold} into \eqref{eq:cost_LS} and present the MSE cost function,
\begin{align}
   \mathcal{E}_{\rm MSE} = \sum\limits_{l\in\mathcal{N}\setminus n}\mathbb{E}\|{\bm B}_{n,l}{\bm x}_{l,t}\|_2^2+\mathbb{E}\|\left({\bm I}_K\otimes {\bm b}_{n}\right)^{\rm H}{\bm v}_t\|_2^2.\label{eq:int_plus_noise}
\end{align}
With the transmission power of the individual active user in each cluster $l$ denoted as $\sigma_l^2$, user activity probability $\alpha_l$ and noise power $\sigma_{v}^2$, \eqref{eq:int_plus_noise} can be simplified as,
\begin{align}
   \mathcal{E}_{\rm MSE} &={\sum\nolimits_{l\in\mathcal{N}\setminus n}\alpha_l\sigma_l^2\|{\bm B}_{n,l}\|_2^2+\sigma_{v}^2\|\left({\bm I}_K\otimes {\bm b}_{n}\right)^{\rm H}\|_2^2}.\label{eq:int_slimplified}
\end{align}
With $\|{\bm B}_{n,l}\|_2^2={\bm b}_n^{\rm H}\sum\nolimits_{k=1}^{K}\tilde{\bm G}_{l,k}\tilde{\bm G}_{l,k}^{\rm H}{\bm b}_n$, we formulate the beamforming optimisation problem as,
\begin{align}
   {\rm arg}\min_{{\bm b}_{n}}\ &{\bm b}_n^{\rm H}\left(\sum\nolimits_{l\in\mathcal{N}\setminus n}\alpha_l\sigma_l^2\sum\nolimits_{k=1}^{K}\tilde{\bm G}_{l,k}\tilde{\bm G}_{l,k}^{\rm H}+K\sigma_{v}^2{\bm I}_M\right){\bm b}_n,\notag\\ {\rm s.t.}\ &{\bm b}_{n}^{\rm H}\bar{\bm a}_n=1.\label{eq:beamforming problem}
\end{align}

Eq. \eqref{eq:beamforming problem} describes a constrained quadratic convex optimisation problem, and the closed-form solution of it for each cluster $n$ is given by,
\begin{align}
   {\bm b}_n^{\rm SBF} = \dfrac{\left(\sum\limits_{l\in\mathcal{N}\setminus n}\alpha_l\sigma_l^2\sum_{k=1}^{K}\tilde{\bm G}_{l,k}\tilde{\bm G}_{l,k}^{\rm H}+K\sigma_{v}^2{\bm I}_M\right)^{-1}\bar{\bm a}_n}{{\bar{\bm a}_n^{\rm H}}\left(\sum\limits_{l\in\mathcal{N}\setminus n}\alpha_l\sigma_l^2\sum_{k=1}^{K}\tilde{\bm G}_{l,k}\tilde{\bm G}_{l,k}^{\rm H}+K\sigma_{v}^2{\bm I}_M\right)^{-1}\bar{\bm a}_n}.\label{eq:SBF}
\end{align}
$K\sigma_{v}^2$ denotes the total noise power, involving the suppression of the additive noise by beamforming. It also acts as a diagonal loading factor to enable the matrix inversion in \eqref{eq:SBF}. $\alpha_l\sigma_l^2$ involves the suppression of the interference signals. Notably, the balance between noise and interference suppression hinges on the interplay between the signal-to-noise ratio (SNR) $\delta_{l}\triangleq \sigma_l^2/\sigma_{v}^2$ and $\alpha_l$, relating to the interfering clusters $l\in\mathcal{N}\setminus n$. Thus, we can pragmatically select an empirical SNR (ESNR) $\delta_{l}$ and a rough $\alpha_l$ from the interval $(0,1]$ without requiring precise values. The solution \eqref{eq:SBF} is referred to as statistical beamforming (SBF), capable of effectively curbing interference even when the number of antenna elements significantly falls short of the number of users.

In practical mMTC scenarios, the small data sample per user is insufficient to represent the statistics in \eqref{eq:beamforming problem} by using the sample variance. In addition, the inaccurate ESNRs and user activity probabilities also influence the tradeoff between the interference and noise suppression to some extent. Thus, it is better to use the LS cost function rather than the MSE.
\subsection{Dynamic Beamforming Scheme}
We now develop the beamforming scheme based on the LS criterion. In light of Eqs. \eqref{eq:BF_received_k}-\eqref{eq:B_nn}, the LS error function in \eqref{eq:cost_LS} can be further expanded as follows,
\begin{align}
\mathcal{E}_{\rm LS}\left({\bm b}_{n},\cdot\right)=\sum\nolimits_{k=1}^K\sum\nolimits_{t=1}^\mathcal{T} \|{\bm b}_n^{\rm H}{\bm y}_{k,t} -{\bm b}_n^{\rm H}\tilde{\bm G}_{n,k}{\bm x}_{n,t}\|_2^2.
\end{align}
Thus, the LS-based beamforming optimisation problem can be further expressed as
\begin{align}
   {\rm arg}\min_{{\bm b}_{n}}\ &\mathcal{E}_{\rm LS}\left({\bm b}_{n},\cdot\right)={\bm b}_n^{\rm H}\sum\nolimits_{k=1}^K\sum\nolimits_{t=1}^\mathcal{T} {\bm i}_{n,k,t}{\bm i}_{n,k,t}^{{\rm H}}{\bm b}_n,\notag\\
    {\rm s.t.}\ &{\bm b}_{n}^{\rm H}\bar{\bm a}_n=1,\label{eq:LS-problem}
\end{align}
where ${\bm i}_{n,k,t}$ is the interference plus the noise component (IpNC), defined as,
\begin{align}
{\bm i}_{n,k,t}\triangleq {\bm y}_{k,t} -\tilde{\bm G}_{n,k}{\bm x}_{n,t}.\label{eq:real IpNC}
\end{align}
Similar to the SBF, the dynamic beamforming (DBF) solution to \eqref{eq:LS-problem} is derived, i.e.,
\begin{equation}
   {\bm b}_n^{\rm DBF} = \left({\bm R}_{n}+\epsilon {\bm I}_M\right)^{-1}\bar{\bm a}_n/\left({\bar{\bm a}_n^{\rm H}}\left({\bm R}_{n}+\epsilon {\bm I}_M\right)^{-1}\bar{\bm a}_n\right)\label{eq:DBF}
\end{equation}
where ${\bm R}_{n}\triangleq 1/(K\mathcal{T})\sum\nolimits_{k=1}^K\sum\nolimits_{t=1}^\mathcal{T} {\bm i}_{n,k,t}{\bm i}_{n,k,t}^{{\rm H}}$  can be seen as the auto-correlation matrix \footnote{In fact, the matrix ${\bm R}_{n}$ is a rough time-average approximation of the auto-correlation matrix due to the small number of slots. Thus, we still refer to the dynamic beamforming herein as a least-squares solution.} of the IpNC, and $\epsilon$ is a diagonal loading factor.

The measurement signal ${\bm y}_{k,t}$ and the transmitted signal ${\bm x}_{n,t}$ are not prerequisites for SBF. Likewise, DBF does not demand prior knowledge of equivalent channel matrices from interfering user clusters. The SBF and DBF approaches are readily applicable to prevailing receive beamforming scenarios, particularly for receivers featuring a limited number of antennas. The DBF simplifies to the conventional constrained least squares (LS) beamforming method when dealing with only one desired user and one subcarrier \cite{Van1988}.

\section{The Integration of Beamforming and Compressed Sensing}\label{sec:algorithm}
The DBF algorithm necessitates prior knowledge of ${\bm x}_{n,t}$ for $n=1,2,\cdots,N$ and $t=1,2,\cdots,\mathcal{T}$, which paradoxically are the signals under estimation. Consequently, we turn our attention towards the joint optimisation of signal estimation and beamforming.

In light of \eqref{eq:received_vector}, the received signal over a frame can be represented in matrix form by, 
\begin{align}
{\bm Y}=\sum\nolimits_{n=1}^N\tilde{\bm G}_{n}{\bm X}_{n}+{\bm V}\in\mathbb{C} ^{KM\times \mathcal{T} },\label{eq:received_matrix}
\end{align}
where the $t$th column vector of ${\bm X}_{n}$ is ${\bm x}_{n,t}$ and the $t$th column of ${\bm V}$ is ${\bm v}_{t}$. 
Similarly, extending ${\bm y}_n$ in \eqref{eq:BF_received} in one frame yields,
\begin{align}
   {\bm Y}_n&=({\bm I}_K\otimes {\bm b}_n)^{\rm H}{\bm Y}\notag\\
   &= {\bm B}_{n,n}{\bm X}_{n}+\sum\nolimits_{l\in\mathcal{N}\setminus n}{\bm B}_{n,l}{\bm X}_{l}+\left({\bm I}_K\otimes {\bm b}_{n}\right)^{\rm H}{\bm V}.\label{eq:BF_received_mat}
\end{align}
To utilise the block sparsity, i.e., constant user activity in a frame, \eqref{eq:BF_received_mat} is vectorised as,
\begin{align}
   {\bm \eta}_n = {\bm {\mathcal{D}}}_{n}{\bm c}_{n} + {\bm z}_n, \label{eq:received vectorisation}
\end{align}   
where ${\bm \eta}_n={\rm vec}\{{\bm Y}_n^{\rm T}\}$, ${\bm {\mathcal{D}}}_n={\bm B}_{n,n}\otimes {\bm I}_{\mathcal{T}}\in\mathbb{C} ^{K\mathcal{T}\times Q\mathcal{T}}$ and ${\bm c}_n={\rm vec}\{{\bm X}_n^{\rm T}\}$. ${\bm z}_n$ is regarded as the IpNC under beamforming. 
Therefore, the joint optimisation problem for any cluster $n$ is rewritten as,
\begin{align}
   {\rm arg}\min_{{\bm b}_{n},{\bm c}_{n}}\ &\mathcal{E}_{\rm LS} \left({\bm b}_{n},{\bm c}_{n}\right)=\|{\bm \eta}_n -{\bm {\mathcal{D}}}_{n}{\bm c}_{n}\|_2^2,\label{eq:joint cost}\\
      {\rm s.t.}\ & \tilde{\Gamma}_{n,1}=\tilde{\Gamma}_{n,2}=\cdots=\tilde{\Gamma}_{n,\mathcal{T} }=\tilde{\Gamma}_n,\notag\\
      \ &|\tilde{\Gamma}_n|\leq \bar{s},\notag\\
      \ & {\bm b}_{n}^{\rm H}\bar{\bm a}_n=1.\notag
\end{align}
For simplicity, we define $\varepsilon_n \triangleq \|{\bm \eta}_n -{\bm {\mathcal{D}}}_{n}{\bm c}_{n}\|_2^2$ as the residual energy of cluster $n$ in the following sections.
\subsection{General Framework for the Joint Optimisation}
\begin{figure}[!t]
   \centerline{\includegraphics[width=0.35\textwidth]{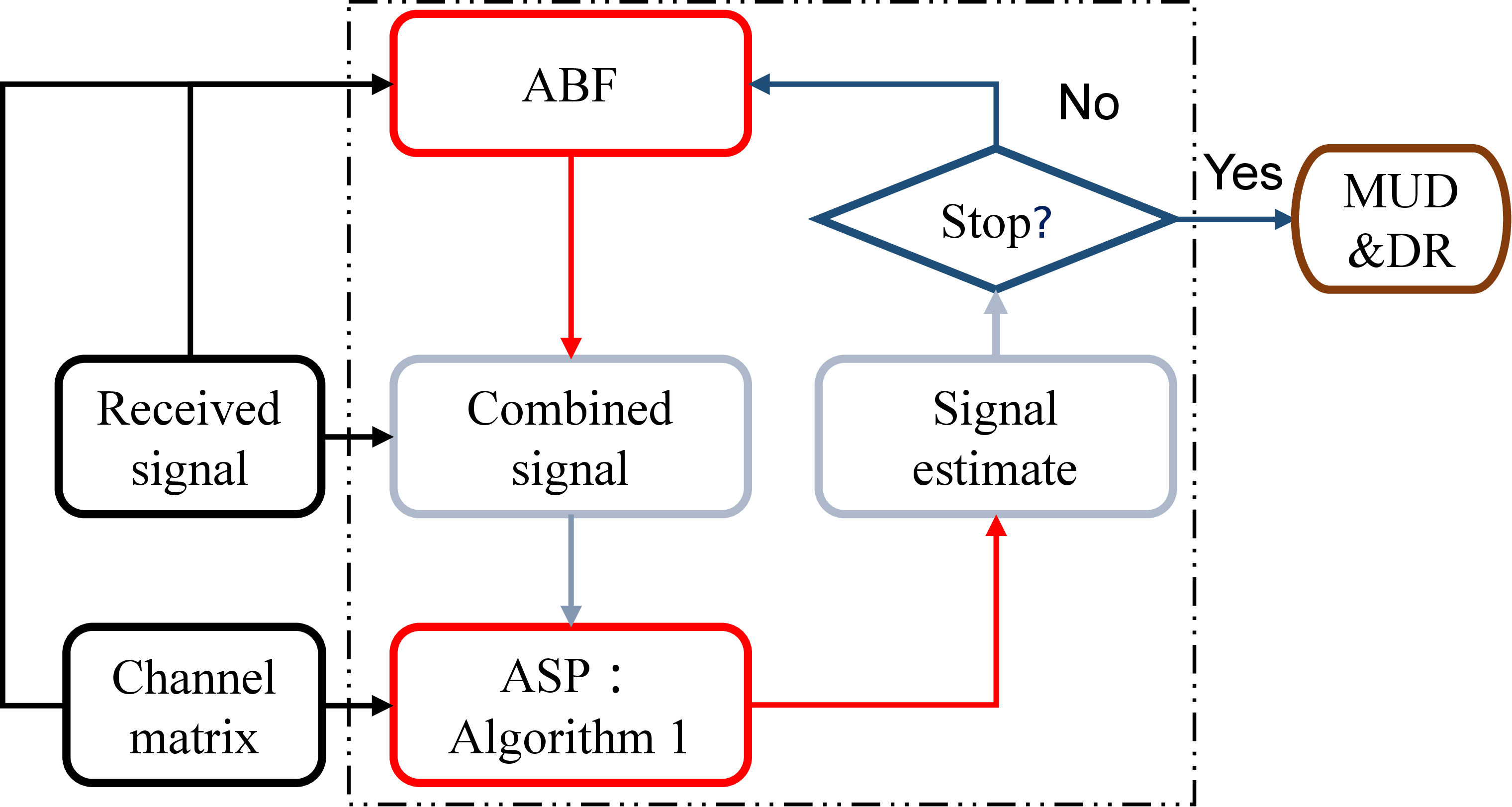}}
   \caption{A general framework of the integration of SDMA and CS-based grant-free NOMA}\label{fig:JABFMP}
   \vspace{-0.3cm}
\end{figure}

As mentioned in Section \ref{sec: related}, CS-based methods can be employed for MUD, such as CoSaMP \cite{Needell2009} and SP \cite{Dai2009,Du2018} \footnote{Other existing multiple user detection methods can also be extended and applied to this framework.}. Before delving into the specifics, we will provide a brief overview of the design principles behind the joint optimisation system. For any cluster $n$, given the known beamforming weight and user sparsity level, the sparse signal recovery problem \eqref{eq:joint cost} can be efficiently solved using CS methods. Subsequently, the signal estimate is used to update the adaptive beamforming (ABF) module, generating new measurements for the CS module. Fig. \ref{fig:JABFMP} illustrates a general framework that integrates SDMA and CS for uplink grant-free access for any user cluster $n$. In this paper, we focus on the block-sparsity based adaptive SP (ASP) method in the CS module.

\subsection{Algorithm Design for the Joint Adaptive Beamforming and Subspace Pursuit}
Based on the beamforming weight $\hat{\bm b}_n$ which is initialised by the SBF weight ${\bm b}_n^{\rm SBF}$ before the first iteration, the measurements (combined signals) for the ASP are generated by,
\begin{equation}
\begin{cases}
\hat{\bm Y}_n=({\bm I}_K\otimes \hat{\bm b}_n)^{\rm H}{\bm Y},\\
\hat{\bm \eta}_n={\rm vec}\{\hat{\bm Y}_n^{\rm T}\}.
\end{cases}\label{eq:measurement update}
\end{equation}
We also have,
\begin{equation}
   \begin{cases}
\hat{\bm B}_{n,n}=({\bm I}_K\otimes \hat{\bm b}_n)^{\rm H}\tilde{\bm G}_n,\\
\hat{\bm {\mathcal{D}}}_n=\hat{\bm B}_{n,n}\otimes {\bm I}_{\mathcal{T}}.
\end{cases}\label{eq:Parameter matrix}
\end{equation}
With the measurement $\hat{\bm \eta}_n$ and the parameter matrix $\hat{\bm {\mathcal{D}}}_n$ of cluster $n$, we can use the ASP algorithm in Algorithm \ref{alg:ASP} to estimate the user support set and the transmitted signals. The finding function $\mathcal{F} (\mathcal{V},\zeta )$ in Algorithm \ref{alg:ASP} selects the indices of the first $\zeta$ largest elements of an ordered set/vector $\mathcal{V}$.

The main steps in Algorithm \ref{alg:ASP} are detailed as follows:\\
\textbf{Step 3:} To estimate the support set $\Lambda$ by adding the current selected $s$ users with larger residual energy into the previously estimated support set $\hat\varGamma_{n,\iota}$.\\
\textbf{Step 4:} To compute the initial signal estimates ${\bm w}[q,\mathcal{T}]$ for all the candidate users in the support set of Step 3.\\
\textbf{Step 5:} To estimate the support set $\hat\varGamma_{n,\iota+1}$ by sparsity level $s$ by selecting the first $s$ largest values of the $l_2$ norms (magnitudes) of ${\bm w}[q,\mathcal{T}]$ over all users in one cluster.\\
\textbf{Step 6:} With the support set estimate $\hat\varGamma_{n,\iota+1}$ at the $\iota$th iteration, the signal is estimated by,
\begin{align}
   \begin{cases}
   \hat{\bm c}_{n, \iota}[\hat\varGamma_{n,\iota+1},\mathcal{T}] = (\hat{\bm {\mathcal{D}}}_n[\hat\varGamma_{n,\iota+1},\mathcal{T}])^\dagger \hat{{\bm \eta}}_n,\\
   \hat{\bm c}_{n, \iota}[\mathcal{Q}\setminus \hat\varGamma_{n,\iota+1},\mathcal{T}]=0,
\end{cases}\label{eq:signal estimate}
\end{align} 
where $\mathcal{Q}$ is the set of user indices for any cluster. \\
We denote the vector ${\bm c}_n[q,\mathcal{T}]$ as the $q$th $\mathcal{T}\times 1$ vector block of ${\bm c}_n$ and the matrix ${\bm {\mathcal{D}}}_{n}[q,\mathcal{T}]$ as the matrix block of ${\bm {\mathcal{D}}}_{n}$ constituted by consecutive columns with index from $(q-1)\mathcal{T}+1$ to $q\mathcal{T}$. Furthermore, ${\bm c}_n[\Lambda,\mathcal{T}]$ and ${\bm {\mathcal{D}}}_{n}[\Lambda,\mathcal{T}]$ denote the sub-vector and sub-matrix by selecting their respective blocks according to the indices from the set $\Lambda$.

Subsequently, with the output $\hat{\bm X}_n=[{\rm vec}^{-1}(\hat{\bm c}_n,\mathcal{T})]^{\rm T}$ of the ASP, the IpNC is estimated by,
\begin{align}
\hat{\bm i}_{n,k,t}={\bm y}_{k,t} -\tilde{\bm G}_{n,k}\hat{\bm x}_{n,t},\label{eq:IpNC estimate}
\end{align}
 with $\hat{\bm x}_{n,t}$ being the $t$th column of $\hat{\bm X}_n$. The beamforming weight is accordingly updated by,
\begin{align}
   \hat{\bm b}_n = \left(\hat{\bm R}_{n}+\epsilon {\bm I}_M\right)^{-1}\bar{\bm a}_n/\left({\bar{\bm a}_n^{\rm H}}\left(\hat{\bm R}_{n}+\epsilon {\bm I}_M\right)^{-1}\bar{\bm a}_n\right),\label{eq:Adaptive DBF}
\end{align}
 with the estimation of the auto-correlation matrix ${\bm R}_{n}$,
 \begin{align}
   \hat{\bm R}_{n}\triangleq 1/(K\mathcal{T})\sum\nolimits_{k=1}^K\sum\nolimits_{t=1}^\mathcal{T} \hat{\bm i}_{n,k,t}\hat{\bm i}_{n,k,t}^{{\rm H}}.\label{eq:est_Rn}
 \end{align}

\begin{algorithm}[!t]
   \footnotesize
	\renewcommand{\algorithmicrequire}{\textbf{Input:}}
	\renewcommand{\algorithmicensure}{\textbf{Output:}}
	\caption{Adaptive subspace pursuit algorithm}
	\label{alg:ASP}
   \textbf{Input:} The measurement signal $\hat{\bm \eta}_n$, the parameter matrix $\hat{\bm {\mathcal{D}}}_n$, the initial support set $\hat\varGamma_{n,1}$, the initial residual $\hat{\bm r}_{n,1}$ and the maximum iteration number $\mathcal{L}_1$. \\
   \textbf{Output:} Signal estimation $\hat{\bm c}_{n,\iota-2}$, active user set $\hat\varGamma_{n,\iota-1}$ and residual $\hat{\bm r}_{n,\iota-1}$.
	\begin{algorithmic}[1]
      \STATE Initial iteration index $\iota=1$,
      \REPEAT
      \STATE (Support estimation) $\Lambda =\hat\varGamma_{n,\iota}\cup \mathcal{F} (\{\|\hat{\bm {\mathcal{D}}}_n^{\rm H}[q,\mathcal{T}]\hat{\bm r}_{n,\iota}\|_2^2\}_\mathcal{Q},s)$.
      \STATE (LS estimation) ${\bm w}[\Lambda,\mathcal{T}] = (\hat{\bm {\mathcal{D}}}_n[\Lambda,\mathcal{T}])^\dagger \hat{{\bm \eta}}_n$, ${\bm w}[\mathcal{Q}\setminus {\Lambda},\mathcal{T}]=0$.
      \STATE (Support pruning) $\hat\varGamma_{n,\iota+1} =\mathcal{F} (\{\|{\bm w}[q,\mathcal{T}]\|_2^2\}_\mathcal{Q},s)$.
      \STATE (Signal estimation) $\hat{\bm c}_{n, \iota}[\hat\varGamma_{n,\iota+1},\mathcal{T}] = (\hat{\bm {\mathcal{D}}}_n[\hat\varGamma_{n,\iota+1},\mathcal{T}])^\dagger \hat{{\bm \eta}}_n$, $\hat{\bm c}_{n, \iota}[\mathcal{Q}\setminus \hat\varGamma_{n,\iota+1},\mathcal{T}]=0$.
      \STATE (Residual update) $\hat{\bm r}_{n,\iota+1} = \hat{{\bm \eta}}_n-\hat{\bm {\mathcal{D}}}_n\hat{\bm c}_{n, \iota}$, $\iota=\iota+1$.
      \UNTIL {$\|\hat{\bm r}_{n,\iota}\|_2^2\geq\|\hat{\bm r}_{n,\iota-1}\|_2^2$} or $\iota-1=\mathcal{L}_1$.
	\end{algorithmic}
\end{algorithm}
\begin{algorithm}[!t]
   \footnotesize
	\renewcommand{\algorithmicrequire}{\textbf{Input:}}
	\renewcommand{\algorithmicensure}{\textbf{Output:}}
	\caption{Joint adaptive beamforming and subspace pursuit algorithm: user detection}
	\label{alg:J-ABF-SP}
   \textbf{Input:} The received signals ${\bm Y}$, equivalent channel matrices $\tilde{\bm G}_n$,
   number of time slots $\mathcal{T}$, upper bound for user sparsity level $\bar{s}$, SBF weight ${\bm b}_n^{\rm SBF}$ in \eqref{eq:SBF}, diagonal loading factor $\epsilon$, stopping factor $\vartheta_1$, average steering vector $\bar{\bm a}_n$, and the maximum iteration $\mathcal{L}_1$ for user detection. \\
   \textbf{Output:} Reconstructed sparse signal ${\bm X}_{n,1}$, support set $\tilde{\varGamma}_n$ and residual energy $e_{n}$  for each $n\in \mathcal{N}$
	\begin{algorithmic}[1]
      \FOR{each cluster $n\in \mathcal{N}$}
      \STATE (Support initialisation) Null initial support set $\varGamma_{0}=\varnothing$.
      \STATE (Measurement initialisation) Compute ${{\bm \eta}}_n$ and ${\bm {\mathcal{D}}}_n$ via \eqref{eq:measurement update} and \eqref{eq:Parameter matrix} by using ${\bm b}_n^{\rm SBF}$ to replace $\hat{\bm b}_n$.\\
      \FOR{sparsity $s=1$ to $\bar{s}$}
      \STATE (Measurement initialisation) The iterative index $z =1$, $\hat{{\bm \eta}}_n={{\bm \eta}}_n$ and $\hat{\bm {\mathcal{D}}}_n={\bm {\mathcal{D}}}_n$.\\
      \STATE (Residual and support initialisation) $\hat{\bm r}_{z}=\hat{\bm \eta}_n$ and $\hat\varGamma_{z}=\varGamma_{s-1}$.\\
      \REPEAT
      \STATE (Residual and support initialisation) $\hat{\bm r}_{n,1}=\hat{\bm r}_{z}$, $\hat\varGamma_{n,1}=\hat\varGamma_{z}$.\\
      \STATE Invoking the ASP algorithm.
      \STATE (Parameter passing) $z=z+1$, $\hat{\bm c}_{z}=\hat{\bm c}_{n,\iota-2}$, $\hat\varGamma_{z}=\hat\varGamma_{n,\iota-1}$ and $\hat{\bm r}_{z}=\hat{\bm r}_{n,\iota-1}$.
      \STATE (Beamforming weight) $\hat{\bm X}_n=[{\rm vec}^{-1}(\hat{\bm c}_{z},\mathcal{T})]^{\rm T}$, compute $\hat{\bm i}_{n,k,t}$ by \eqref{eq:IpNC estimate}, and compute $\hat{\bm b}_{n,z}$ by \eqref{eq:Adaptive DBF}.
      \STATE (Measurement update) Compute $\hat{{\bm \eta}}_n$ and $\hat{\bm {\mathcal{D}}}_n$ via \eqref{eq:measurement update} and \eqref{eq:Parameter matrix} by using $\hat{\bm b}_{n,z}$ to replace $\hat{\bm b}_n$.
      \UNTIL $|\|\hat{\bm r}_{z}\|_2^2-\|\hat{\bm r}_{z-1}\|_2^2|/\|\hat{\bm r}_{z-1}\|_2^2<\vartheta_1$
      \STATE (Sparsity update) ${\bm c}_s=\hat{\bm c}_{z-1}$, $\varepsilon_s=\|\hat{\bm r}_{z-1}\|_2^2$ and $\varGamma_{s} = \hat\varGamma_{z-1}$.
      \STATE (TPR update) $\hat{\bm X}_n=[{\rm vec}^{-1}({\bm c}_s,\mathcal{T})]^{\rm T}$, compute $\hat\gamma_{n,s}$ by \eqref{eq:range_s}.
      \ENDFOR
      \STATE (Candidate sparsity set) $\mathcal{S}_c  = \mathcal{S}\setminus\{s\in\mathcal{S}:\hat\gamma_{n,s}>\acute{\gamma}_{n}\}$.\\
      \STATE (Sparsity decision) $s_o={\rm arg}\min\nolimits_{s\in\mathcal{S}_c}{\varepsilon_s}$,
      \STATE (Active user set) $\tilde{\varGamma}_n = \varGamma_{s_o}$.
      \STATE (Residual energy) $e_{n} = \varepsilon_{s_o}$.
      \STATE (Signal recovery) ${\bm X}_{n,1}=[{\rm vec}^{-1}({\bm c}_{s_o},\mathcal{T})]^{\rm T}$.
      \ENDFOR
	\end{algorithmic}
\end{algorithm}

To sum up, a joint adaptive beamforming and subspace pursuit algorithm (J-ABF-SP) is presented in Algorithm \ref{alg:J-ABF-SP}. Considering the potential small fluctuation of the sparsity level due to the empirical user activity rate $\alpha_l$, the upper bound $\bar{s}$ for sparsity level searching is selected within a range, e.g., $(\alpha_lQ, 2\alpha_lQ]$. We now detail the main steps of Algorithm \ref{alg:J-ABF-SP}.

\textbf{Parallel computation:} The iteration process (the steps between 2 and 21) can be performed in parallel for all clusters in $\mathcal{N}$. This guarantees the fairness in terms of the access delay for different user clusters and thus reduces the total latency in comparison to the serial computation.

\textbf{Parameter passing:} The outputs of ASP encompass the estimate of the support set (active user set), residual, and signal estimate (step 9), with the latter employed for beamforming updates (step 11). The updated beamforming weight contributes to generating new measurements (step 12). These, along with the support set and residual, are then fed back into ASP (steps 8 and 9). Upon fulfilling the stopping condition of adaptive beamforming (step 13), the signal estimates, residual energy, and support set estimate are preserved (step 14). Notably, only the support set estimate proceeds to the next iteration at a fresh sparsity level (step 6). These parameter passing processes ensure the continuity of the entire iteration.

\textbf{Important initialisation:} 
We initialise the beamforming for each sparsity level using the SBF weight (step 3). The SBF offers effective channel utilisation for both the desired user cluster and the interfering user clusters, even without precise SNR values. However, the adaptive beamforming weight at the current sparsity level cannot be directly applied in the next sparsity level iteration. This is because the beamformer treats the signals of undetected active users (UDAUs) as interferences (steps 5 and 12) when the given sparsity is smaller than the actual sparsity level. This aspect is explained in more detail in Appendix \ref{sec:Appen_residual}. Consequently, the residual at each sparsity level is initialised using the measurement vector generated through the SBF weight (step 6).

\textbf{Stopping condition:} For the ASP (step 9), the stopping condition is that the current residual energy (norm) is larger than the previous one (step 8 in Algorithm \ref{alg:ASP}), which indicates the current and subsequent iterations tend to deteriorate the user detection and signal recovery performance. For the beamforming update (step 13), we employ a threshold related to the change in residual energy as the stopping criterion. This helps prevent unnecessary beamforming updates.

\subsection{Error Analysis}\label{sec:error analysis}
We now analyse the signal estimation error when using the J-ABF-SP algorithm. The combined signal \eqref{eq:received vectorisation} for cluster $n$ is expressed in a sparse matrix form, i.e.,
\begin{align}
   {\bm \eta}_n = {\bm {\mathcal{D}}}_{n}[\tilde{\varGamma}_n,\mathcal{T}]{\bm c}_{n}[\tilde{\varGamma}_n,\mathcal{T}] + {\bm z}_n, \label{eq:received sparse}
\end{align}
where $\tilde{\varGamma}_n$ is the index set of the active users in cluster $n$ and ${\bm z}_n$ is the IpNC under beamforming. With the support set estimate $\varGamma_s$, the transmitted signals are estimated via \eqref{eq:signal estimate}, i.e.,
\begin{equation}
   \hat{\bm c}_n[\varGamma_s,\mathcal{T}] = ({\bm {\mathcal{D}}}_n[\varGamma_s,\mathcal{T}])^\dagger ({\bm {\mathcal{D}}}_{n}[\tilde{\varGamma}_n,\mathcal{T}]{\bm c}_{n}[\tilde{\varGamma}_n,\mathcal{T}] + {\bm z}_n).\label{eq:signal estimate sparse}
\end{equation} 
Considering that ${\bm {\mathcal{D}}}_n[\varGamma_s,\mathcal{T}]$ is with the full column rank, we have
\begin{equation} 
({\bm {\mathcal{D}}}_n[\varGamma_s,\mathcal{T}])^\dagger=\left(({\bm {\mathcal{D}}}_n[\varGamma_s,\mathcal{T}])^{\rm H}{\bm {\mathcal{D}}}_n[\varGamma_s,\mathcal{T}]\right)^{-1}({\bm {\mathcal{D}}}_n[\varGamma_s,\mathcal{T}])^{\rm H}.
\end{equation}
Thus, we have $({\bm {\mathcal{D}}}_n[\varGamma_s,\mathcal{T}])^\dagger{\bm {\mathcal{D}}}_{n}[\varGamma_s,\mathcal{T}]={\bm I}$.

We now simplify \eqref{eq:signal estimate sparse} as,
\begin{align}
   \hat{\bm c}_n[\varGamma_s,\mathcal{T}] =\ &\big[{\bm {\mathcal{D}}}_n[\varGamma_{n,s},\mathcal{T}],{\bm {\mathcal{D}}}_n[\varGamma_s\setminus \varGamma_{n,s},\mathcal{T}]\big]^\dagger \notag\\
   &\cdot({\bm {\mathcal{D}}}_{n}[\tilde{\varGamma}_n,\mathcal{T}]{\bm c}_{n}[\tilde{\varGamma}_n,\mathcal{T}] + {\bm z}_n).\label{eq:signal estimate simplify}
\end{align}
where $\varGamma_{n,s}=\tilde{\varGamma}_n\cap \varGamma_s$ denotes the index set of the detected
active users (DAUs). We have $\varGamma_{n,s}\subseteq  \varGamma_s$ and $\varGamma_{n,s}\subseteq \tilde{\varGamma}_n$. In the following, we will analyse the signal estimation error under
two cases, i.e., no falsely detected inactive users (FDIUs)
exists with $\varGamma_{n,s}= \varGamma_s$ and FDIUs exist with $\varGamma_{n,s}\subset \varGamma_s$.

Firstly, for $\varGamma_{n,s} = \varGamma_s$, we have $\varGamma_s\setminus \varGamma_{n,s}=\emptyset$ and the signal estimates of DAUs in \eqref{eq:signal estimate simplify} can be rewritten as
\begin{align}
   \hat{\bm c}_n[\varGamma_s,\mathcal{T}] =\ &({\bm {\mathcal{D}}}_n[\varGamma_s,\mathcal{T}])^\dagger \bigg({\bm {\mathcal{D}}}_{n}[\varGamma_s,\mathcal{T}]{\bm c}_{n}[\varGamma_s,\mathcal{T}]\notag\\
    &+ {\bm {\mathcal{D}}}_{n}[\tilde{\varGamma}_n\setminus \varGamma_s,\mathcal{T}]{\bm c}_{n}[\tilde{\varGamma}_n\setminus \varGamma_s,\mathcal{T}] + {\bm z}_n\bigg)\notag\\
    =\ &{\bm c}_{n}[\varGamma_s,\mathcal{T}]+ ({\bm {\mathcal{D}}}_n[\varGamma_s,\mathcal{T}])^\dagger\notag\\
    &\cdot({\bm {\mathcal{D}}}_{n}[\tilde{\varGamma}_n\setminus \varGamma_s,\mathcal{T}]{\bm c}_{n}[\tilde{\varGamma}_n\setminus \varGamma_s,\mathcal{T}] + {\bm z}_n).\label{eq:signal estimate missed}
\end{align}
If $\varGamma_s\subset \tilde{\varGamma}_n$, we can find that the signal estimates of DAUs are contaminated by the received signals from the UDAUs and the IpNC simultaneously. The existence of the UDAUs indicates the information loss. When $\varGamma_s=\tilde{\varGamma}_n$, there is no UDAU and \eqref{eq:signal estimate missed} can be simplified as,
\begin{align}
   \hat{\bm c}_n[\tilde{\varGamma}_n,\mathcal{T}] = {\bm c}_{n}[\tilde{\varGamma}_n,\mathcal{T}]+({\bm {\mathcal{D}}}_n[\tilde{\varGamma}_n,\mathcal{T}])^\dagger {\bm z}_n. \label{eq:signal estimate normal}
\end{align}
It can be seen that more accurate signal estimates are generated in \eqref{eq:signal estimate normal} than those in \eqref{eq:signal estimate missed} since they are impacted solely by
the IpNC.

Secondly, when FDIUs exist with $\varGamma_{n,s} \subset \varGamma_s$, \eqref{eq:signal estimate simplify} can be
rewritten as, 
\begin{equation}
   \begin{aligned}
      \hat{\bm c}_n[\varGamma_s,\mathcal{T}] =\ &\begin{bmatrix}({\bm {\mathcal{D}}}_n[\varGamma_{n,s},\mathcal{T}])^\dagger-{\bm F}_{n,s}\\{\bm W}_{n,s}^{\rm H}\end{bmatrix}\big({\bm {\mathcal{D}}}_{n}[\varGamma_{n,s},\mathcal{T}]{\bm c}_{n}[\varGamma_{n,s},\mathcal{T}]\\
      & + {\bm {\mathcal{D}}}_{n}[\tilde{\varGamma}_n\setminus \varGamma_{n,s},\mathcal{T}]{\bm c}_{n}[\tilde{\varGamma}_n\setminus \varGamma_{n,s},\mathcal{T}] + {\bm z}_n\big)\label{eq:signal estimate alarm_missed}
   \end{aligned}
   \end{equation}
   where 
   \begin{align}
      \begin{cases}
         {\bm F}_{n,s} = ({\bm {\mathcal{D}}}_n[{\varGamma}_{n,s},\mathcal{T}])^\dagger{\bm {\mathcal{D}}}_n[\varGamma_s\setminus {\varGamma}_{n,s},\mathcal{T}]{\bm W}_{n,s}^{\rm H},\\
         {\bm W}_{n,s} = {\bm U}_{n,s}({\bm U}_{n,s}^{\rm H}{\bm U}_{n,s})^{-1},\\
         {\bm U}_{n,s} = ({\bm I}-{\bm {\mathcal{D}}}_n[{\varGamma}_{n,s},\mathcal{T}]({\bm {\mathcal{D}}}_n[{\varGamma}_{n,s},\mathcal{T}])^\dagger){\bm {\mathcal{D}}}_n[\varGamma_s\setminus {\varGamma}_{n,s},\mathcal{T}].
      \end{cases}
   \end{align}
   Note that the relevant matrix inversion can be referred to Appendix \ref{sec:Appen_MP_inv}.
   Based on the property ${\bm F}_{n,s}{\bm {\mathcal{D}}}_{n}[{\varGamma}_{n,s},\mathcal{T}]={\bm W}_{n,s}^{\rm H}{\bm {\mathcal{D}}}_{n}[{\varGamma}_{n,s},\mathcal{T}]={\bm 0}$, we have from \eqref{eq:signal estimate alarm_missed},
   \begin{align}
      \hat{\bm c}_n[{\varGamma}_{n,s},\mathcal{T}] =\ &{\bm c}_{n}[{\varGamma}_{n,s},\mathcal{T}]+(({\bm {\mathcal{D}}}_n[{\varGamma}_{n,s},\mathcal{T}])^\dagger-{\bm F}_{n,s})\notag\\
      &\big( {\bm {\mathcal{D}}}_{n}[\tilde{\varGamma}_n\setminus \varGamma_{n,s},\mathcal{T}]{\bm c}_{n}[\tilde{\varGamma}_n\setminus \varGamma_{n,s},\mathcal{T}] + {\bm z}_n\big)\notag\\
      =\ &{\bm c}_{n}[{\varGamma}_{n,s},\mathcal{T}]+({\bm {\mathcal{D}}}_n[{\varGamma}_{n,s},\mathcal{T}])^\dagger\notag\\
      &\cdot({\bm I}-{\bm {\mathcal{D}}}_n[\varGamma_s\setminus {\varGamma}_{n,s},\mathcal{T}]{\bm W}_{n,s}^{\rm H})\notag\\
      &\cdot\big( {\bm {\mathcal{D}}}_{n}[\tilde{\varGamma}_n\setminus \varGamma_{n,s},\mathcal{T}]{\bm c}_{n}[\tilde{\varGamma}_n\setminus \varGamma_{n,s},\mathcal{T}] + {\bm z}_n\big),\label{eq:detected_active signal estimate}
   \end{align}
   and 
   \begin{align}
      \hat{\bm c}_n[\varGamma_s\setminus \varGamma_{n,s},\mathcal{T}] =\ &{\bm W}_{n,s}^{\rm H}{\bm {\mathcal{D}}}_{n}[\tilde{\varGamma}_n\setminus \varGamma_{n,s},\mathcal{T}]{\bm c}_{n}[\tilde{\varGamma}_n\setminus \varGamma_{n,s},\mathcal{T}]\notag\\
      & + {\bm W}_{n,s}^{\rm H}{\bm z}_n. \label{eq:falsely-detected signal estimate}
   \end{align}

On one hand, when UDAUs exist with $\varGamma_{n,s}\subset\tilde{\varGamma}_n$, the signal estimates $\hat{\bm c}_n[{\varGamma}_{n,s},\mathcal{T}]$ for DAUs in \eqref{eq:detected_active signal estimate} face contamination from both received signals emanating from UDAUs and IpNC, similar to \eqref{eq:signal estimate missed} with $\varGamma_{s}\subset\tilde{\varGamma}_n$. However, due to the unit non-zero eigenvalues of ${\bm {\mathcal{D}}}_n[\varGamma_s\setminus {\varGamma}_{n,s},\mathcal{T}]{\bm W}_{n,s}^{\rm H}$, the overall interference power, stemming from both the UDAUs and IpNC, is anticipated to be lower than that in \eqref{eq:signal estimate missed}. This results in more precise signal estimates.

The signal estimates for FDIUs in \eqref{eq:falsely-detected signal estimate} encompass contributions from both received signals from the UDAUs and
IpNC, weighted by ${\bm W}^{\rm H}_{n,s}$, different from \eqref{eq:inactive signal estimate} with $\varGamma_{n,s}=\tilde{\varGamma}_n$. Specifically, signal estimate magnitudes for FDIUs typically fall short of those attributed to DAUs in \eqref{eq:detected_active signal estimate}. The degradation of the magnitudes is due to the channel differences between various users, as exemplified by ${\bm W}_{n,s}^{\rm H}{\bm {\mathcal{D}}}_{n}[\tilde{\varGamma}_n\setminus \varGamma_{n,s},\mathcal{T}]$ in \eqref{eq:falsely-detected signal estimate}, where ${\bm W}_{n,s}$ involves the channels of DAUs and FDIUs while ${\bm {\mathcal{D}}}_{n}[\tilde{\varGamma}_n\setminus \varGamma_{n,s},\mathcal{T}]$ involves the channels of UDAUs.

On the other hand, when all active users are detected with $\varGamma_{n,s}=\tilde{\varGamma}_n$, we have the signal estimates as follows, in light of \eqref{eq:detected_active signal estimate} and \eqref{eq:falsely-detected signal estimate},
\begin{align}
   \hat{\bm c}_n[\tilde{\varGamma}_n,\mathcal{T}] =\ &{\bm c}_{n}[\tilde{\varGamma}_n,\mathcal{T}]+({\bm {\mathcal{D}}}_n[\tilde{\varGamma}_n,\mathcal{T}])^\dagger\notag\\
   &({\bm I}-{\bm {\mathcal{D}}}_n[\varGamma_s\setminus \tilde{\varGamma}_n,\mathcal{T}]{\bm W}^{\rm H}){\bm z}_n,\label{eq:active users signal}\\
   \hat{\bm c}_n[\varGamma_s\setminus \tilde{\varGamma}_n,\mathcal{T}] =\ &{\bm W}^{\rm H}{\bm z}_n. \label{eq:inactive signal estimate}
\end{align}
where 
\begin{align}
   \begin{cases}
      {\bm W} = {\bm U}({\bm U}^{\rm H}{\bm U})^{-1},\\
      {\bm U} = {\bm {\mathcal{D}}}_n[\varGamma_s\setminus \tilde{\varGamma}_n,\mathcal{T}]\\
      ~~~~~~-{\bm {\mathcal{D}}}_n[\tilde{\varGamma}_n,\mathcal{T}]({\bm {\mathcal{D}}}_n[\tilde{\varGamma}_n,\mathcal{T}])^\dagger{\bm {\mathcal{D}}}_n[\varGamma_s\setminus \tilde{\varGamma}_n,\mathcal{T}].
   \end{cases} \label{eq:matrix component}
\end{align}
It can be seen that the signal estimates $\hat{\bm c}_n[\tilde{\varGamma}_n,\mathcal{T}]$ of the active users  suffer from the additive IpNC weighted by $({\bm {\mathcal{D}}}_n[\tilde{\varGamma}_n,\mathcal{T}])^\dagger({\bm I}-{\bm {\mathcal{D}}}_n[\varGamma_s\setminus \tilde{\varGamma}_n,\mathcal{T}]{\bm W}^{\rm H})$ while the signal estimates for FDIUs are constituted by the IpNC weighted by ${\bm W}^{\rm H}$. Since ${\bm {\mathcal{D}}}_n[\varGamma_s\setminus \tilde{\varGamma}_n,\mathcal{T}]{\bm W}^{\rm H}$ has unit non-zero eigenvalues as ${\bm W}^{\rm H}{\bm {\mathcal{D}}}_n[\varGamma_s\setminus \tilde{\varGamma}_n,\mathcal{T}]={\bm I}$, \eqref{eq:active users signal} is subject to relatively minor interference from the IpNC and may yield more accurate signal estimates than those from \eqref{eq:signal estimate normal}. Nonetheless, the $\varGamma_s\supset\tilde{\varGamma}_n$ scenario inevitably leads to false alarms.

For simplicity, we have considered the same beamforming weight for the above analysis, indicating the same IpNC
under beamforming. In fact, as detailed in Appendix \ref{sec:Appen_residual}, the beamforming weight varies in different sparsity levels, leading to distinct IpNCs under beamforming.
\subsection{Sparsity Level Decision}
Expectantly, the accurate support set estimate $\varGamma_s$ satisfies $\varGamma_s=\tilde{\varGamma}_n$ with $s$ equal to the actual sparsity level $s_o$. We now study the sparsity level decision method via the signal estimate $\hat{\bm c}_n[\varGamma_s,\mathcal{T}]$ above. 

Define the temporal power ratio (TPR) as,
\begin{align} 
   \gamma_n \triangleq  \dfrac{\max_{q\in\tilde{\varGamma}_n}{\|{\bm x}_{n,q}\|_2^2}}{\min_{q\in\tilde{\varGamma}_n}{\|{\bm x}_{n,q}\|_2^2}},
\end{align}
where ${\bm x}_{n,q}$, the transmitted signal vector of the user $u_{n,q}$ in one sampling duration, is the transpose of the $q$th row of the transmitted signal matrix ${\bm X}_{n}$. Similarly, TPR of $\hat{\bm x}_{n,q}$ with given sparsity level $s$ is defined as,
\begin{align} 
   \hat{\gamma}_{n,s} \triangleq  \dfrac{\max_{q\in\varGamma_s}{\|\hat{\bm x}_{n,q}\|_2^2}}{\min_{q\in\varGamma_s}{\|\hat{\bm x}_{n,q}\|_2^2}},\label{eq:range_s}
\end{align}
where $\hat{\bm x}_{n,q}=\hat{\bm c}_n[q,\mathcal{T}]$ is a block vector of the above signal estimate $\hat{\bm c}_n[\varGamma_s,\mathcal{T}]$.

The error analysis in Section \ref{sec:error analysis} motivates us to develop a user sparsity level decision method, i.e., \\
1) The candidate sparsity set $\mathcal{S}_c = \mathcal{S}\setminus\{s\in\mathcal{S}:\hat\gamma_{n,s}>\acute{\gamma}_{n}\}$ with $\mathcal{S}=\{1,2,\cdots,\bar{s}\}$.\\
2) The sparsity is given by $s_o={\rm arg}\min\limits_{s\in\mathcal{S}_c}{\varepsilon_s}$.\\
We analyse the feasibility of this method in the following.

The TPR within a given sampling duration $\mathcal{T}$ generally remains below a specific threshold. In particular, when $\mathcal{T}$ is suitably large, the temporal power of the transmitted signal approaches its actual transmission power. Assuming uniform transmission power among active users within the same cluster \footnote{Should active users within the same cluster exhibit different transmission power, the TPR will approach the maximum transmission power ratio among them.}, $\gamma_n$ tends to converge towards 1. As inferred from \eqref{eq:signal estimate missed}, \eqref{eq:signal estimate normal}, \eqref{eq:detected_active signal estimate} and \eqref{eq:active users signal}, the signal estimates of DAUs are affected by the IpNC and may even be adversely influenced by UDAUs. In contrast, the TPR is a relative metric and is less susceptible to such concerns. Considering the influence of randomness due to limited samples, it is reasonable to empirically set a threshold $\acute{\gamma}_n$ greater than 1.

As discussed in \eqref{eq:inactive signal estimate}, if inactive users are mistakenly identified as active, their signal estimates are dominated by the IpNC, which is notably suppressed by beamforming. This results in $\hat{\gamma}_{n,s}>\acute{\gamma}_n$. Even when UDAUs and FDIUs coexist with $\varGamma_{n,s}\subset  \varGamma_s$ and $\varGamma_{n,s}\subset \tilde{\varGamma}_n$, the signal estimate magnitudes of FDIUs in (37) are generally lower than those of DAUs in (36). Consequently, step 1) is employed to eliminate sparsity levels where FDIUs probably exist. Step 2) aims to ascertain the user sparsity level via the fact that the residual energy decreases as the sparsity level $s$ approaches the true value. This verification is presented in Appendix \ref{sec:Appen_residual}.

\subsection{Interference Cancellation}
As analysed earlier, the transmitted signal is estimated by \eqref{eq:signal estimate} via the measurements generated by beamforming for the received signal in \eqref{eq:measurement update}. However, the IpNC suppression solely relying on beamforming may be limited, especially with the number of antennas comparable to the number of user clusters. We propose an interference cancellation (IC) scheme to further improve the signal estimation based on the support set and
initial signal estimates from the J-ABF-SP algorithm.
\begin{figure}[!t]
   \centerline{\includegraphics[width=0.4\textwidth]{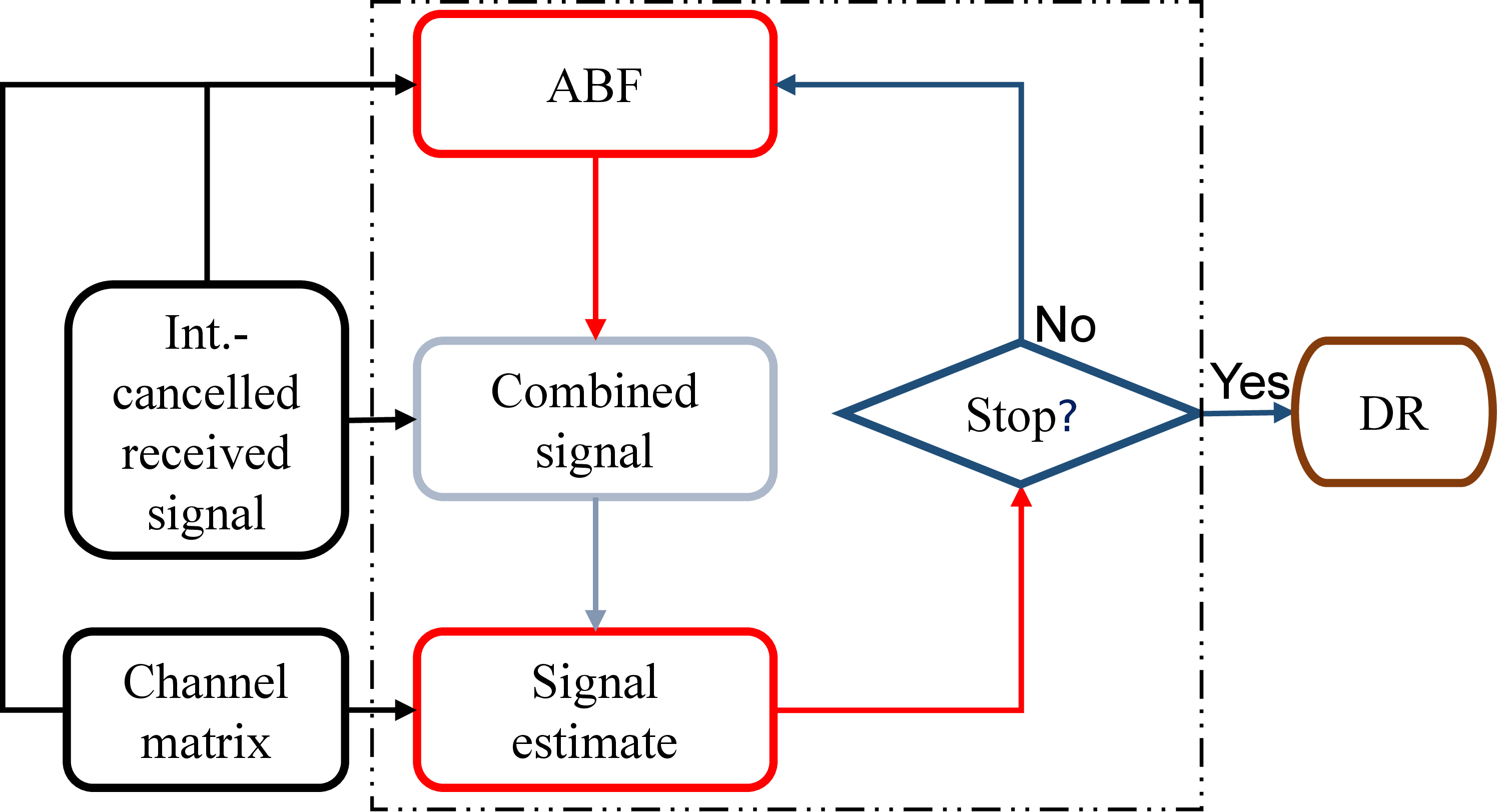}}
   \caption{The main flowchart for Algorithm 3}\label{fig:IC_flowChart}
\end{figure}
\begin{algorithm}[!t]
   \footnotesize
   \renewcommand{\algorithmicrequire}{\textbf{Input:}}
   \renewcommand{\algorithmicensure}{\textbf{Output:}}
   \caption{Interference cancellation enhanced signal recovery}
   \label{alg:J-ABF-SP-IC}
   \textbf{Input:} The received signals ${\bm Y}$, equivalent channel matrices $\tilde{\bm G}_n$,
   number of the consecutive time slots $\mathcal{T}$, diagonal loading factor $\epsilon$, average steering vector $\bar{\bm a}_n$, maximum number of iterations $\mathcal{L}_2$ and $\mathcal{L}_3$, active user set $\tilde{\varGamma}_n$, initial error $e_{n}$ and initial signal estiamtion ${\bm X}_{n,1}$. \\
   \textbf{Output:} Reconstructed sparse signal ${\bm X}_{n}$
   \begin{algorithmic}[1]
      \STATE (Weight initialisation) For each cluster $n$, $\hat{\bm X}_{n}={\bm X}_{n,1}$, $\hat{\bm i}_{n,k,t}={\bm y}_{k,t}-\tilde{\bm G}_{n,k}\hat{\bm x}_{n,t}$, compute $\hat{\bm b}_n$ by \eqref{eq:Adaptive DBF}.
      \STATE (Error initialisation) For each cluster $n$, $\tilde{e}_{1,n}=e_{n}$.
      \FOR{Iteration $\iota_2=1$ to $\mathcal{L}_2$}
      \FOR{Cluster $n=1$ to $N$}
      \STATE (Interference reconstruction) construct the received interference signal ${\bm Y}_{i,n}=\sum\nolimits_{l=1,l\neq n}^N\tilde{\bm G}_{l}{\bm X}_{l,\iota_2}$.
      \STATE (Interference cancellation) ${\mathcal{\bm Y}}_n = {\bm Y}-{\bm Y}_{i,n}$.
      \FOR{Iteration $\iota_3=1$ to $\mathcal{L}_3$}
      \STATE (Measurement update) Compute $\hat{{\bm \eta}}_n$ and $\hat{\bm {\mathcal{D}}}_n$ using $\hat{\bm b}_n$ via \eqref{eq:measurement update_IC} and \eqref{eq:Parameter matrix}.\\
      \STATE (Signal estimation) $\hat{\bm c}_n[\tilde{\varGamma}_n,\mathcal{T}] = (\hat{\bm {\mathcal{D}}}_n[\tilde{\varGamma}_n,\mathcal{T}])^\dagger\hat{\bm \eta}_n$, $\hat{\bm c}_n[\mathcal{Q}\setminus \tilde{\varGamma}_n,\mathcal{T}]=0$.\\
      \STATE (Residual update) $\tilde{e}_{\iota_3+1,n}=\|\hat{\bm \eta}_n-\hat{\bm {\mathcal{D}}}_n\hat{\bm c}_n\|_2^2$.
      \IF{$\tilde{e}_{\iota_3+1,n}<\tilde{e}_{\iota_3,n}$ and $\iota_3<\mathcal{L}_3$}
      \STATE (Beamforming weight) $\hat{\bm X}_n=[{\rm vec}^{-1}(\hat{\bm c}_n,\mathcal{T})]^{\rm T}$, $\hat{\bm i}_{n,k,t}={\bm y}_{k,t} -\tilde{\bm G}_{n,k}\hat{\bm x}_{n,t}$, and compute $\hat{\bm b}_n$ by \eqref{eq:Adaptive DBF}.
      \ELSE
      \STATE (Residual modification) $\tilde{e}_{1,n}=\tilde{e}_{\iota_3,n}$.
      \STATE Break.
      \ENDIF
      \ENDFOR
      \STATE (Signal update) ${\bm X}_{n,\iota_2+1}=\hat{\bm X}_n$.
      \ENDFOR
      \ENDFOR
      \STATE (Signal recovery) ${\bm X}_n={\bm X}_{n,\mathcal{L}_2+1}$.
   \end{algorithmic}
\end{algorithm}

With the active user set and initial signal estimates from the J-ABF-SP algorithm, we can reconstruct the received signal from each cluster $n$ as
$\tilde{\bm G}_{n}{\bm X}_{n,\iota}$,
where ${\bm X}_{n,\iota}$ is the signal estimate after the $(\iota-1)$th IC. Then, we can obtain the IC-enabling received signal for cluster $n$, i.e.,
\begin{align}
   {\mathcal{\bm Y}}_n = {\bm Y}-{\bm Y}_{i,n},\label{eq:IC}
\end{align}
where ${\bm Y}_{i,n}=\sum\nolimits_{l=1,l\neq n}^N\tilde{\bm G}_{l}{\bm X}_{l,\iota}$ is the reconstructed interference signal for cluster $n$. Then, the new measurements are generated by, 
\begin{equation}
   \begin{cases}
   \hat{\bm Y}_n=({\bm I}_K\otimes \hat{\bm b}_n)^{\rm H}{\mathcal{\bm Y}}_n,\\
   \hat{\bm \eta}_n={\rm vec}\{\hat{\bm Y}_n^{\rm T}\},
   \end{cases}\label{eq:measurement update_IC}
\end{equation}
Note that $\hat{\bm b}_n$ is computed by \eqref{eq:Adaptive DBF} based on the signal estimate $\hat{\bm X}_{n}$, which is initialised by ${\bm X}_{n,1}$ before the first IC. In addition, the parameter matrix $\hat{\bm {\mathcal{D}}}_n$ is computed by \eqref{eq:Parameter matrix}. Based on the measurements \eqref{eq:measurement update_IC}, the transmitted signals can be estimated by using \eqref{eq:signal estimate}.

The detailed steps on IC-enhanced signal recovery are summarised in Algorithm \ref{alg:J-ABF-SP-IC}, which mainly consists of three loops. Loop 1 gives the number $\mathcal{L}_2$ to perform the IC which is generally small since the performance enhancement by \eqref{eq:IC} typically reaches its peak quickly. The steps in loop 2 can be performed in parallel for all clusters. This parallel computation property, similar to Algorithm \ref{alg:J-ABF-SP}, ensures fairness among different user clusters in terms of access delay and computational resources. Loop 3 is used to iterate the signal estimation and beamforming based on the constructed interference-cancelled received signal, with major procedures outlined in Fig. \ref{fig:IC_flowChart}. Similar to the ASP algorithm, the stopping condition for loop 3 is that the current residual energy is larger than the previous one. The residual energy, signal estimate, and beamforming weight in loop 3 will be conveyed to loop 1 as initial values. The algorithms \ref{alg:J-ABF-SP} and \ref{alg:J-ABF-SP-IC} are referred to as the IC-enhanced joint adaptive beamforming and subspace pursuit algorithm (J-ABF-SP-IC).
\section{Computational Complexity Analysis}
In this section, we compare the computational complexity of the proposed algorithms with benchmark methods, including TA-BSASP \cite{Du2018}, OAMP-MMV-SSL \cite{Mei2022}, OAMP-MMV-ASL \cite{Mei2022}, and DS-AMP \cite{Qiao2022MM} methods. The complexity is measured by the number of complex-valued multiplications needed for the whole algorithm implementation.  
\begin{table*}[!t]
       \centering
       \footnotesize
       \caption{The number of complex-valued multiplications}
       \vspace{-0.1cm}
      \label{tab:complexity}	
       \begin{tabular}{|l|c|c|}
           \hline
           Algorithm&Number of complex multiplications&O notation\\\hline
           OAMP-MMV-SSL \cite{Mei2022}&$\mathcal{L}_1((3Q+1){\mathcal{T}}K+(\dfrac{13}{4}P+\dfrac{25}{4}){\mathcal{T}}Q)$&$O(\mathcal{L}_1K\mathcal{T}Q)$\\\hline
           OAMP-MMV-ASL \cite{Mei2022}&$\mathcal{L}_1((3Q+1){\mathcal{T}}K+(\dfrac{13}{4}P+\dfrac{27}{4}){\mathcal{T}}Q+3{\mathcal{T}}^2Q)$&$O(\mathcal{L}_1K\mathcal{T}Q)$\\\hline
           TA-BSASP \cite{Du2018}&$\sum\limits_{s=1}^{s_o}\mathcal{C}_{\rm SP}$&$O(\mathcal{L}_1K\mathcal{T}^3s_o^3)$\\\hline
           DS-AMP \cite{Qiao2022MM}&$\mathcal{L}_1NQ\mathcal{T}M_t(5M_D/2+P+1/4)$&$O(\mathcal{L}_1NQ\mathcal{T}M_t(M_D+P))$\\\hline
           J-ABF-SP&${\mathcal{C}}_{\rm MUD}$&$O(NKQM^2+\mathcal{L}_b\mathcal{L}_1K\mathcal{T}^3\bar{s}^3+\mathcal{L}_b\bar{s}^2(M+K{\mathcal{T}})M^2)$\\\hline
           J-ABF-SP-IC&${\mathcal{C}}_{\rm MUD}+{\mathcal{C}}_{\rm IC}$&$\makecell{O(NKQM(M+\mathcal{L}_2\mathcal{T})+\mathcal{L}_b\mathcal{L}_1K\mathcal{T}^3\bar{s}^3\\
           +(\mathcal{L}_b\bar{s}^2+\mathcal{L}_2\mathcal{L}_3)(M+K{\mathcal{T}})M^2)}$\\\hline
      \end{tabular}
   \vspace{-0.3cm}
\end{table*}

The number of complex-valued multiplications for various algorithms is listed in Table \ref{tab:complexity}. For ease of analysis, we assume the same maximum number of iterations for all methods, i.e., $\mathcal{L}_1$. For the OAMP-MMV-SSL, OAMP-MMV-ASL and DS-AMP, the letter $P$ denotes the dimension of the signal constellation, e.g., $P=2$ for binary phase shift keying (BPSK). For the DS-AMP, $M_D$ is the number of BS antennas, and $M_t=2^{M_{\rm RF}}$ is the number of mirror activation patterns by using media modulation with $M_{\rm RF}$ denoting the number of radio frequency (RF) mirrors.

We now detail the computational complexity of our proposed algorithms for one cluster since the algorithms can be performed in parallel for all clusters. Given the number of alternating iterations as ${\mathcal{L}}_b$, the computational complexity of the J-ABF-SP algorithm is expressed as,
\begin{align}
{\mathcal{C}}_{\rm MUD}=\ &{\mathcal{C}}_{\rm SBF}+MK(Q+{\mathcal{T}})\notag\\
&+{\mathcal{L}}_b\sum\nolimits_{s=1}^{\bar{s}}{\mathcal{C}}_{\rm SP}+\dfrac{\bar{s}(\bar{s}-1)}{2}\mathcal{T}\notag\\
&+\dfrac{{\mathcal{L}}_b\bar{s}(\bar{s}-1)}{2}({\mathcal{C}}_{\rm BF}+MK(Q+{\mathcal{T}})+\mathcal{T}K),
\end{align} 
where ${\mathcal{C}}_{\rm SBF}=M^3+((N-1)KQ+1)M^2+M$ is the complexity for the SBF, ${\mathcal{C}}_{\rm SP}={\mathcal{L}}_1(2Ks^2{\mathcal{T}}^3+2(KQ+Ks){\mathcal{T}}^2+(2Q+K){\mathcal{T}})$ is the complexity for the ASP in Algorithm \ref{alg:ASP}  and ${\mathcal{C}}_{\rm BF}=M^3+(K{\mathcal{T}}+1)M^2+(Q+1)M$ denotes the complexity for beamforming update. 
Given the actual user sparsity level $s_o$, the complexity for the IC-enhanced method in Algorithm \ref{alg:J-ABF-SP-IC} is,
\begin{align}
   {\mathcal{C}}_{\rm IC}=\ &(\mathcal{L}_2\mathcal{L}_3+1){\mathcal{C}}_{\rm BF}+\mathcal{L}_2(N-1)M{\mathcal{T}}KQ\notag\\
   &+\mathcal{L}_2\mathcal{L}_3(Ks_o^2{\mathcal{T}}^3+(s_o+Q)K{\mathcal{T}}^2\notag\\
   &+MK(Q+\mathcal{T})+{\mathcal{T}}K),
\end{align} 
Consequently, the total computational complexity of the J-ABF-SP-IC is ${\mathcal{C}}_{\rm MUD}+{\mathcal{C}}_{\rm IC}$. 

As mentioned in Section \ref{Sec:syst model}, the number of user clusters and the angular distribution range of users within each cluster should match the number of antennas. Therefore, we assume $M$ is in the same magnitude with $N$. Additionally, the signal recovery by the subspace pursuit method requires the number of measurements $K$ no less than $2s_o$ \cite{Dai2009}. Thus, for the J-ABF-SP algorithm, the complexity can be finally denoted by the O notation, 
\begin{align}
   {\mathcal{C}}_{\rm MUD}=O(NKQM^2+\mathcal{L}_b\bar{s}^2(M+K{\mathcal{T}})M^2+\mathcal{L}_b\mathcal{L}_1K\mathcal{T}^3\bar{s}^3),\notag
\end{align}
where the first two terms are directly relevant with the beamforming and the last term is involved with the ASP algorithm.
Similarly, we have the order of the complexity of performing IC, i.e.,
\begin{align}
   {\mathcal{C}}_{\rm IC}=O(\mathcal{L}_2NKQM\mathcal{T}+\mathcal{L}_2\mathcal{L}_3(M+K{\mathcal{T}})M^2+\mathcal{L}_2\mathcal{L}_3Ks_o^2{\mathcal{T}}^3).\notag
\end{align}
Consequently, the total complexity of the J-ABF-SP-IC algorithm is given by 
\begin{align}
   {\mathcal{C}}=\ &O(NKQM(M+\mathcal{L}_2\mathcal{T})+(\mathcal{L}_b\bar{s}^2+\mathcal{L}_2\mathcal{L}_3)     (M+K{\mathcal{T}})M^2\notag\\
   &+\mathcal{L}_b\mathcal{L}_1K\mathcal{T}^3\bar{s}^3).\label{eq:complexity_JABFSPIC}
\end{align}
For ease of analysis, we assume $M=\varsigma+N$ with $\varsigma$ denotes a non-negative integer enabling the number of user clusters $N$ and the angular distribution range of users within each cluster matched to the number of antennas $M$. We further assume $K={Q}/{2}={Q_{all}}/{2N}$ with $Q_{\rm all}=NQ$ denoting the total number of users. Then, \eqref{eq:complexity_JABFSPIC} can be converted into,
\begin{align}
   {\mathcal{C}}=\ &O(Q_{\rm all}^2(M+\mathcal{L}_2\mathcal{T})+(M^3+MQ_{\rm all}{\mathcal{T}})(\mathcal{L}_b\bar{s}^2+\mathcal{L}_2\mathcal{L}_3)\notag\\
   &+\mathcal{L}_b\mathcal{L}_1Q_{\rm all}\mathcal{T}^3\bar{s}^3/(M-\varsigma)).\label{eq:simplified complexity_JABFSPIC}
\end{align}
It is evident that the complexity regarding the number of antennas presents a decreasing-then-increasing trend. It can achieve a minimum value simply by letting the sum of the first two terms equal to the third term in \eqref{eq:simplified complexity_JABFSPIC}. 

In fact,  $\mathcal{L}_b$ denotes the number for beamforming update, which is generally small. For the proposed algorithms, the increased complexity due to beamforming is modest compared to the TA-BSASP algorithms when utilizing a small number of BS antennas. However, the complexity is comparatively high when compared to the OAMP-MMV-SSL and OAMPMMV-ASL methods because they employ complexity reduction schemes, while the proposed algorithms still leverage the the block-sparsity-based ASP method (Algorithm \ref{alg:ASP}) for the MUD.
 
Additionally, it may seem that the proposed algorithms entail higher complexity than the DS-AMP algorithm. However,
the latter relies on a massive number of antennas, whereas our methods can achieve satisfactory performance even with
a small number of antennas, provided that the number of user clusters and the angular distribution range of users within
each cluster match the number of antennas. The complexity of integrating SDMA and grant-free access is expected to
be reduced by using specially designed MUD schemes. This aspect will be investigated in our future work.


\section{Simulation Results}
We now assess the MUD and DR performance of the proposed J-ABF-SP algorithms through simulations. A BS with $M$ antenna elements is considered, serving massive users simultaneously. The users are assumed to be grouped based on the channel correlation into $N\le M$ clusters with $Q$ users in each cluster $n, n=1,2,\cdots,N$. Without loss of generality, we consider $N=3$ and $Q=40$. Assume the AoAs of the users in each cluster are randomly distributed over an angle range with a width of 5 degrees\footnote{As mentioned in Section \ref{Sec:sig model}, the angle range of the clustered users should be generally smaller than the 3 dB beamwidth.}, with the central angles being -30, -10 and 10 degrees, respectively. 

All users employ the common $K=20$ subcarriers, unless specified otherwise. The same spreading signatures, generated in Appendix \ref{sec:Appen_Zadoff}, are utilised in all clusters. In this case, the frequency-domain system overloading factor is $NQ/K=600\%$, which increases linearly with the number of user clusters. We consider the user activity rate to be $\alpha_n=10\%$. Without loss of generality, we consider a typical value $s_o=4$ or $s_o=5$ for the number of active users in each cluster, which is far less than the number of the total users. Each data frame consists of $\mathcal{T} = 7$ continuous symbol durations, following the LTE-Advanced standard \cite{3GPPPCM}.

We consider the detection error rate (DER) and the symbol error rate (SER) as performance metrics. For any cluster $n$, the DER is defined as $p_{d,n} = (f_n+m_n)/Q$ where $f_n$ and $m_n$ denote the number of FDIUs and the number of UDAUs, respectively. The SER is defined as $p_{s,n} = p_{d,n}+S_{e,n}/(Q\mathcal{T})$ where $S_{e,n}$ denotes the number of error symbols of DAUs. Both the DER and SER are calculated over a large number of independent trials. In the following, we consider the same input SNR $\delta_{n}$ for each user cluster $n\in \mathcal{N}$ and present the average values of the DERs or SERs of the $N$ clusters, unless noted otherwise. 

We evaluate the performance of the proposed J-ABF-SP and J-ABF-SP-IC methods for the MUD and DR, in comparison with some benchmark methods, including the Oracle-BSASP \cite{Du2018}, OAMP-MMV-SSL \cite{Mei2022}, the OAMP-MMV-ASL \cite{Mei2022} and the DS-AMP \cite{Qiao2022MM} methods. Without loss of generality, the
transmitted symbols are randomly generated from 16QAM constellation for all the users. In particular, the {Oracle}-BSAMP method is evaluated with known user sparsity levels. The DS-AMP \cite{Qiao2022MM} algorithm relies on the number of BS antennas, and we consider a BS setup with $M_D=150$ antennas for its simulation. The number radio frequency (RF) mirrors is denoted as $M_{\rm RF}$, e.g., $M_{\rm RF}=0$ or $M_{\rm RF}=2$. For the single-antenna benchmark algorithms including the Oracle-BSASP \cite{Du2018}, OAMP-MMV-SSL \cite{Mei2022} and the OAMP-MMV-ASL \cite{Mei2022}, we consider the single-antenna (e.g., the first antenna) reception of any one user cluster, without the interference from the other two clusters. For the proposed algorithms, $\acute{\gamma}_n=3$ is selected as the sparsity decision threshold for each cluster $n$. We also consider ${\rm ESNR} =13$ dB for the SBF, the SNR of 2dB and the number of antennas $M=5$, unless specified otherwise. For clarity, Table \ref{tab:parameter} details the parameter presentation for different figures.
\begin{table}[!t]
       \centering
       \footnotesize
       \caption{The parameters for different simulations}
       \vspace{-0.1cm}
      \label{tab:parameter}	
       \begin{tabular}{|l|c|}
           \hline
           Simulations&Parameters\\\hline
           Figs. \ref{fig:DER_SNR}-\ref{fig:DER_SER_AP}&$s_o=4$\\\hline
           Figs. \ref{fig:DER_ESNR}-\ref{fig:SER_ESNR}&$M=4$ and $s_o=4$\\\hline
           Figs. \ref{fig:SNR_unbalanced}-\ref{fig:Slots_CSIErr}&$s_o=5$\\\hline
      \end{tabular}
   \vspace{-0.3cm}
\end{table}

\begin{figure}[!t]
   \centerline
   {\includegraphics[width=0.38\textwidth]{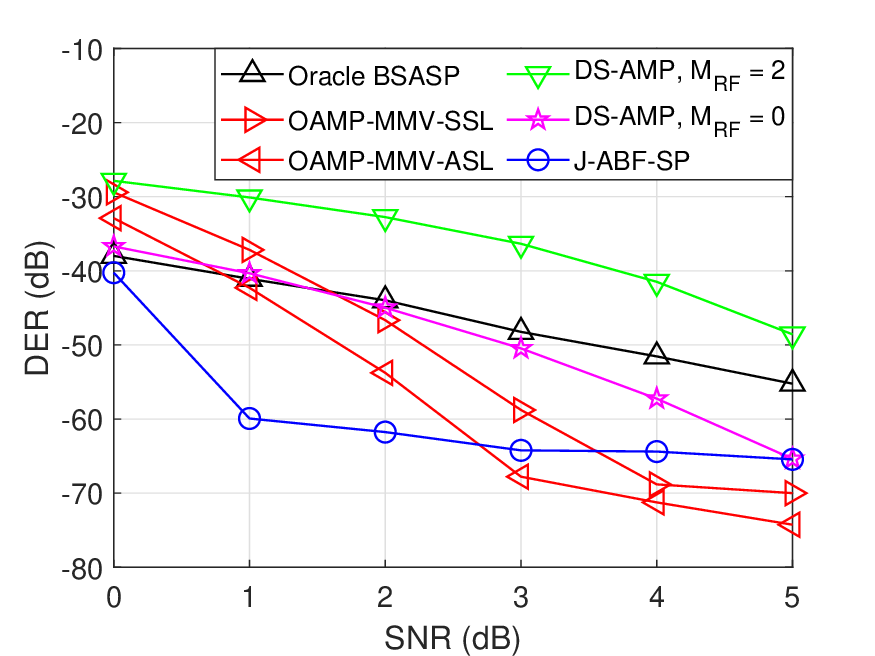}}
   \caption{The DER with respect to SNR}\label{fig:DER_SNR}
   \vspace{-0.3cm}
\end{figure}
\begin{figure}[!t]
   \centerline
   {\includegraphics[width=0.38\textwidth]{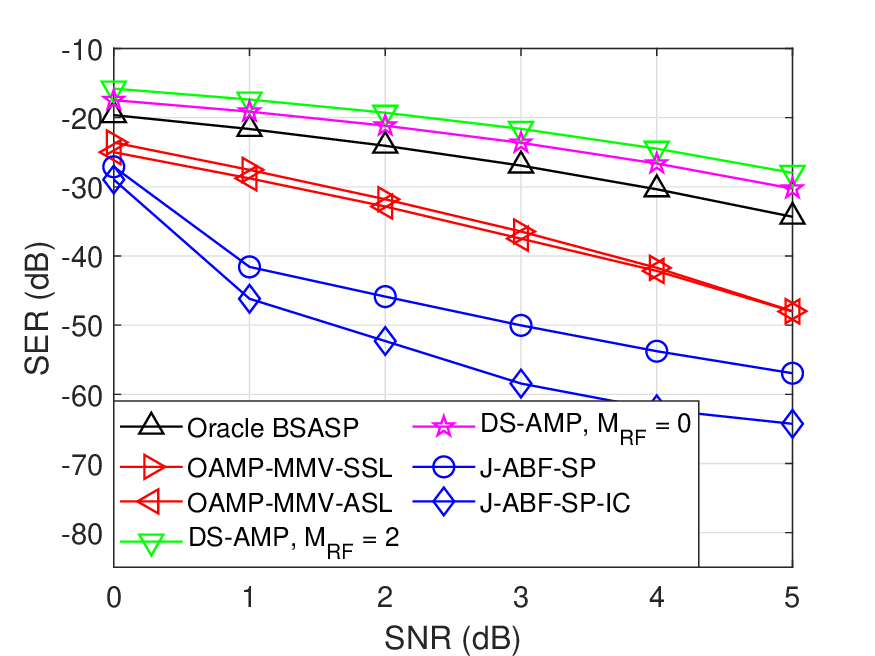}}
   \caption{The SER with respect to SNR}\label{fig:SER_SNR}
   \vspace{-0.3cm}
\end{figure}
Fig. \ref{fig:DER_SNR} shows the DERs regarding the input SNRs for different MUD methods. The proposed J-ABF-SP algorithm performs better in user detection than the Oracle-BSASP algorithm even though the latter knows the user sparsity level a priori. This is because both the SBF and ABF used in the J-ABF-SP can suppress the IpNC contained in the received signal, leading to a higher receiver signal-to-interference-plus-noise ratio (SINR) than that of the Oracle-BSASP. With increasing input SNR for each cluster, the power of corresponding inter-cluster interferences rises uniformly, leading to a SINR (signal-to-interference-plus-noise ratio) floor that induces the DER (detection error rate) floor at a certain input SNR level, e.g., 1 dB. From another perspective, the J-ABF-SP algorithm can achieve extremely low DERs even at low SNRs, e.g., -60 dB DER under the 1 dB SNR. In this regard, it does not matter that the J-ABF-SP presents a slightly higher DER than that of the OAMP-MMV algorithms as the SNR increases to a certain value, e.g., 4 dB. Additionally, the results show that the J-ABF-SP always outperforms the DS-AMP algorithms over the given SNR range.

Figure \ref{fig:SER_SNR} depicts the SERs across various input SNRs. Notably, the proposed J-ABF-SP algorithm showcases a remarkable SER gain of over 8 dB when compared to the OAMPMMV algorithms and exhibits notably superior performance
than other benchmark algorithms. Furthermore, the J-ABFSP-IC algorithm outperforms the J-ABF-SP algorithm. This
improvement can be attributed to IC enhancing the SINR at the receiver.

\begin{figure}[!t]
   \centerline
   {\includegraphics[width=0.38\textwidth]{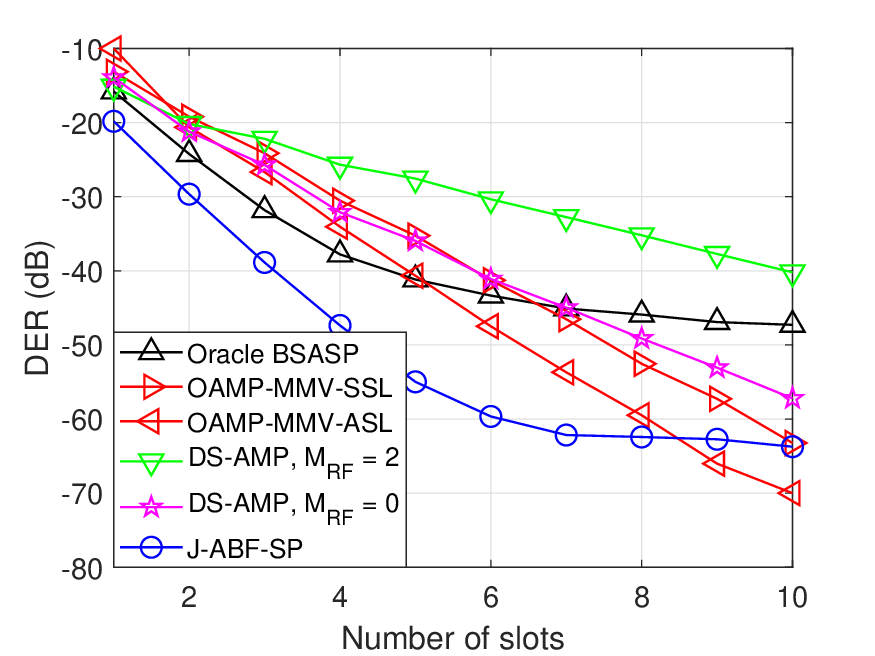}}
   \caption{The DER regarding the number of slots}\label{fig:DER_slots}
   \vspace{-0.3cm}
\end{figure}
\begin{figure}[!t]
   \centerline{\includegraphics[width=0.38\textwidth]{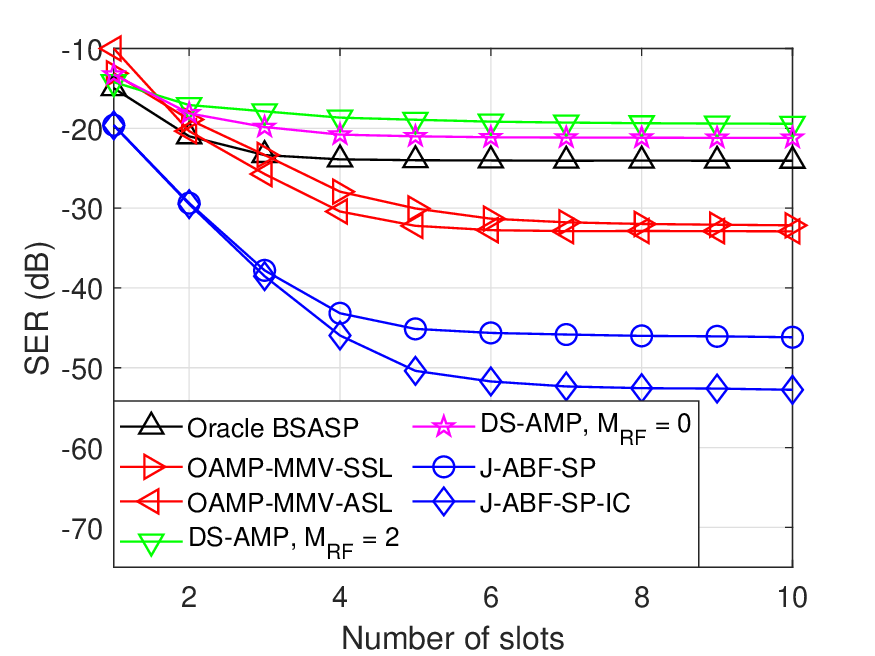}}
   \caption{The SER regarding the number of slots}\label{fig:SER_slots}
   \vspace{-0.3cm}
\end{figure}
Figs. \ref{fig:DER_slots} and \ref{fig:SER_slots} illustrate the DERs and the SERs with respect to the number of slots. The proposed algorithms achieve significantly low DERs and SERs compared to the benchmark algorithms, even with only one slot in a frame. Moreover, the SER performance superiority by the proposed algorithms tends to enhance with the number of slots and eventually converges. In particular, compared with the OAMP-MMV algorithms, the J-ABF-SP algorithm shows slightly inferior DER performance when the number of slots increases to 9, but demonstrates remarkable superiority in SER performance. This indicates that the SER for the DAUs by the proposed algorithms is extremely lower than that of the OAMP-MMV algorithms.

\begin{figure}[!t]
   \centerline
   {\includegraphics[width=0.4\textwidth]{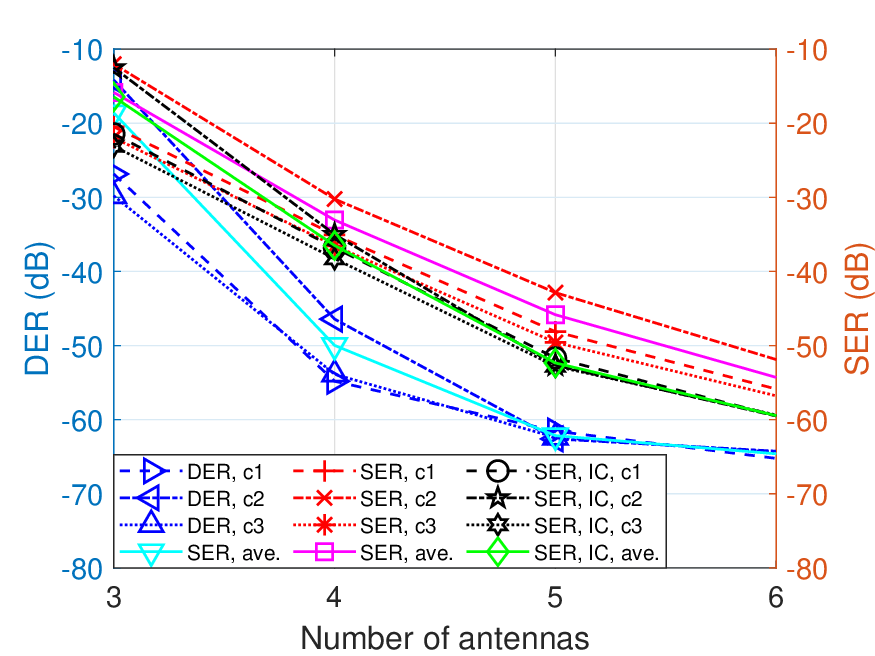}}
   \caption{The DER and SER regarding the number of antennas}\label{fig:DER_SER_AP}
   \vspace{-0.3cm}
\end{figure}

We now study the impact of the number of antennas $M$ on the performance of the proposed algorithms. Fig. \ref{fig:DER_SER_AP} illustrates the DER and SER of each cluster with respect to the number of antennas, respectively. Note that c1 is the abbreviation of cluster 1, similar for c2 and c3, and ave. denotes the average value over three clusters. The DERs of all clusters gradually decrease with the number of antennas. Specifically, the DER of cluster 2 is initially higher than those of the other two clusters with a small number of antennas, but approaches a similar value with the increased number of antennas. This is because cluster 2 is located spatially between the other two clusters and thus suffers from larger interferences, but this impact is mitigated with the enhanced beamforming gain and spatial resolution provided by the increased number of antennas. Similarly, more antennas result in better SERs and smaller SER differences among different clusters. In addition, J-ABF-SP-IC outperforms J-ABF-SP in SER performance. Specifically, the SER performance is enhanced by more than 20 dB by increasing the number of antennas from 4 to 6, indicating a promising prospect for the integration of SDMA and CS for uplink grant-free communication.

We now study the importance of the dynamic update of beamforming weights for the MUD and DR performance. The zero-forcing beamforming (ZFBF) is used as a benchmark \cite{Zhang2019JCN}. We compare the ZFBF-ASP, SBF-ASP, ZFBF-ASP-IC, and SBF-ASP-IC methods, which are obtained by selecting initial beamforming (ZFBF or SBF) and ignoring the beamforming and measurement updates in each iteration in both J-ABF-SP and J-ABF-SP-IC. Specifically, for the SBF-ASP
and SBF-ASP-IC, two ESNRs are considered, i.e., 13 dB or 20 dB. We also consider unbalanced SNRs in distinct clusters,
e.g., SNR=$\{2, 5, 3\}$ in dB for the corresponding clusters $n=\{1, 2, 3\}$, but with the same ESNRs of 13 dB.

\begin{figure}[!t]
   \centerline
   {\includegraphics[width=0.39\textwidth]{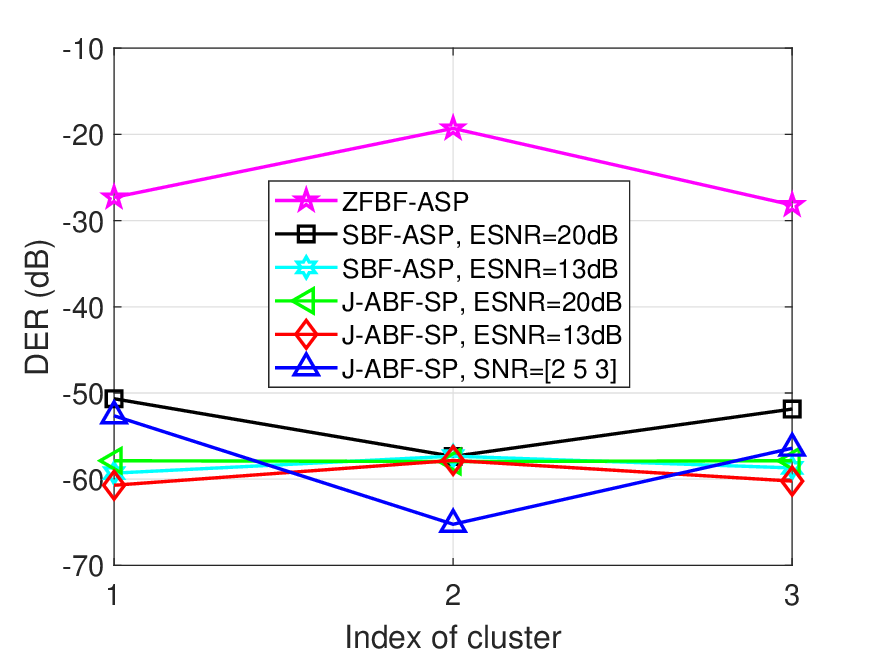}}
   \caption{The DER under different beamforming conditions}\label{fig:DER_ESNR}
   \vspace{-0.3cm}
\end{figure}
\begin{figure}[!t]
   \centerline
   {\includegraphics[width=0.39\textwidth]{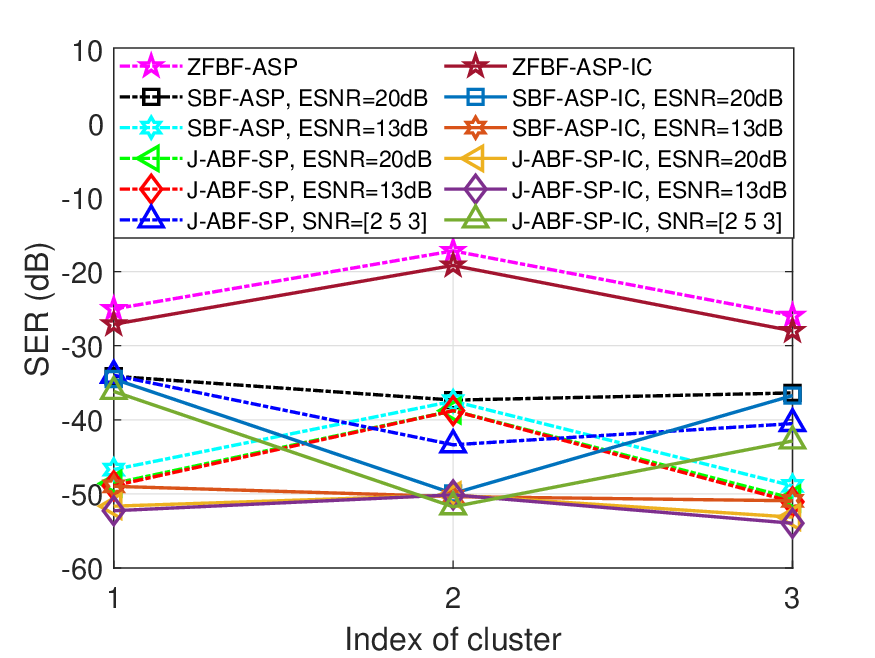}}
   \caption{The SER under different beamforming conditions}\label{fig:SER_ESNR}
   \vspace{-0.3cm}
\end{figure}
Figures \ref{fig:DER_ESNR} and \ref{fig:SER_ESNR} show the DER and SER performance of individual clusters, respectively. We can find that the SBFASP achieves a similar DER or SER with the J-ABF-SP at
ESNR=13 dB, but degraded performance at ESNR=20 dB, while the J-ABF-SP is insensitive to the ESNRs. This indicates
the importance of dynamic beamforming updates when the SNR is unknown a priori. In addition, when compared to the
scenario with the same SNR (5 dB) in all clusters (red line), cluster 2 has a lower DER (SER) while the other two clusters have higher DERs (SERs) in the scenario with different SNRs in different clusters (blue line). This is because the inter-cluster interferences for cluster 2 are weakened since the other two clusters have lower SNRs, while for clusters 1 and 3, their lower SNRs result in their higher DERs (SERs). We also observe from Fig. 12 that the enhanced SER performance can be obtained for all the methods when using IC.

\begin{figure}[!t]
   \centerline
   {\includegraphics[width=0.4\textwidth]{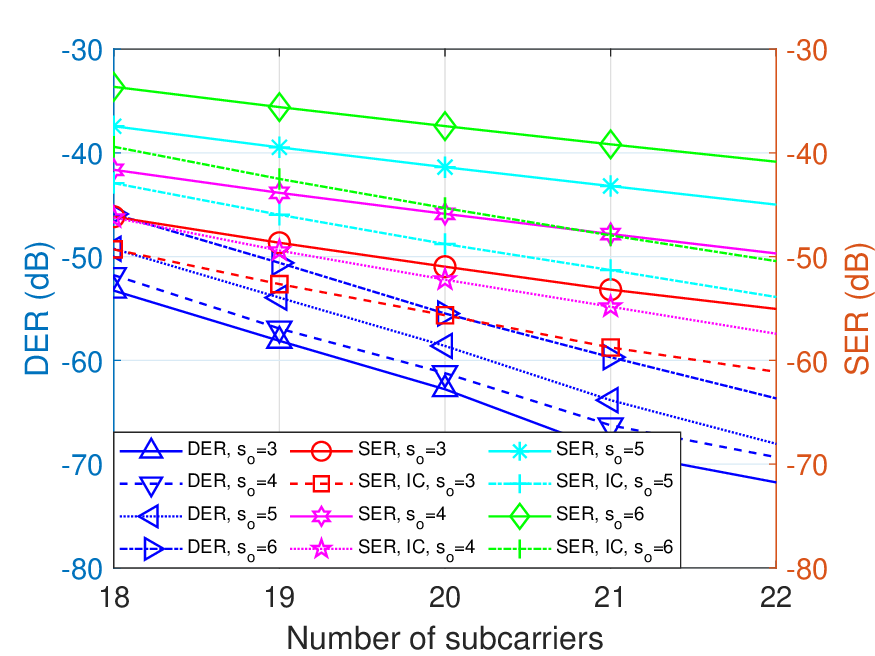}}
   \caption{The DER and SER regarding the number of subcarriers}\label{fig:DER_SER_subc}
   \vspace{-0.3cm}
\end{figure}

The non-orthogonal ZC spreading sequences are considered for simulations, as in Appendix \ref{sec:Appen_Zadoff}. We now study the impact of the number of subcarriers $K$ (length of ZC sequences) on the performance of the proposed algorithms. It is evident from Fig. \ref{fig:DER_SER_subc} that the performance improves gradually with an increasing number of subcarriers, irrespective of their primality. Furthermore, one can also observe the performance enhancement with decreased
user sparsity level $s_o$.

\begin{figure}[!t]
   \centerline
   {\includegraphics[width=0.4\textwidth]{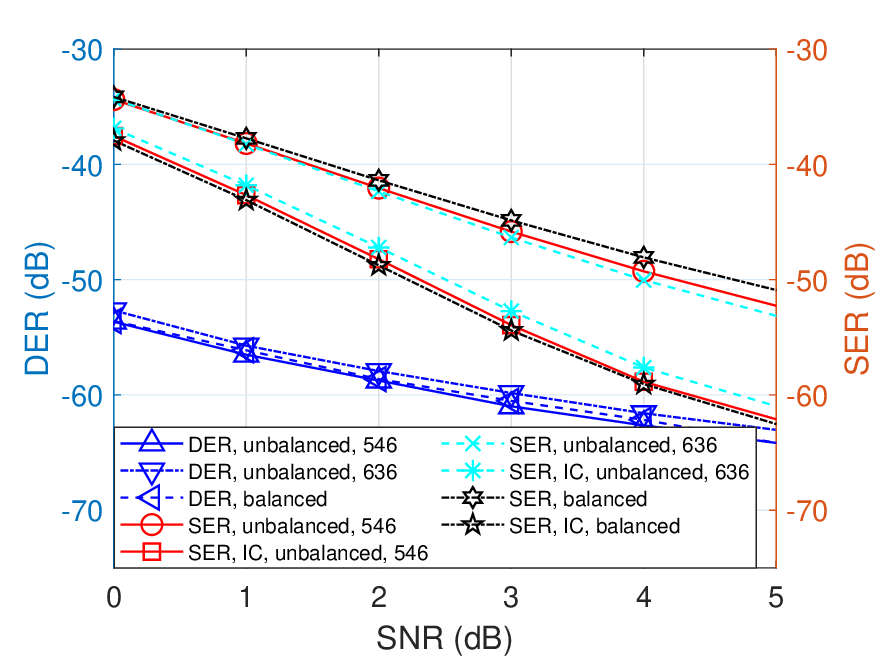}}
   \caption{DER and SER regarding the input SNR with unbalanced clusters}\label{fig:SNR_unbalanced}
   \vspace{-0.3cm}
\end{figure}

\begin{figure}[!t]
   \centerline
   {\includegraphics[width=0.4\textwidth]{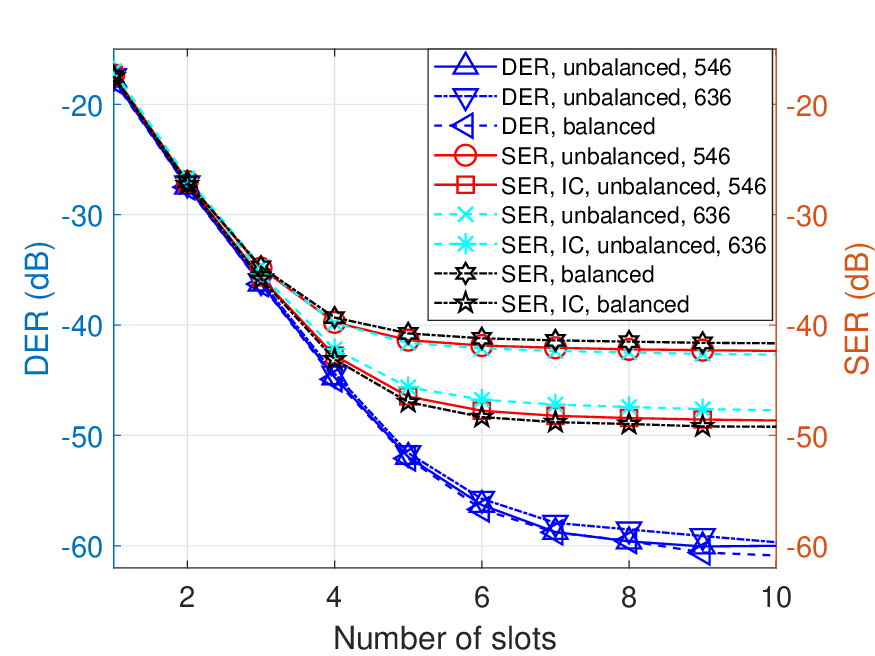}}
   \caption{DER and SER regarding the frame length with unbalanced clusters}\label{fig:Slots_unbalanced}
   \vspace{-0.3cm}
\end{figure}

We now assess the performance of the proposed algorithms in scenarios where clusters have varying numbers of active
users. Without loss of generality, we consider two cases for unbalanced clusters. One is that $s_o = \{5, 4, 6\}$ active users in clusters $n = \{1, 2, 3\}$, respectively. The other is that $s_o = \{6, 3, 6\}$ active users in clusters $n = \{1, 2, 3\}$, respectively. For comparison, we also examine the case with $s_o = 5$ active users in each cluster $n \in \{1, 2, 3\}$. Figs. \ref{fig:SNR_unbalanced} and \ref{fig:Slots_unbalanced} illustrate the DER and SER concerning the input SNR and the number of slots, respectively. We observe that the similar performance can be obtained for both unbalanced and balanced user clusters.

We now explore the effects of the (channel state information) CSI errors on the MUD and DR performance. Assume
there is a random error on the small-scale random fading $\eta_{n,q,k}$, termed as, $\hat\eta_{n,q,k}\sim \mathcal{U} (\eta_{n,q,k}-\delta^{\eta}_{n,q,k}, \eta_{n,q,k}+\delta^{\eta}_{n,q,k})$ with the half disturbation range $\delta^{\eta}_{n,q,k}$. Similarly, assume a random error on each element of the steering vector $a_{n,q,m}$ in channel measurement, termed as, $\hat{a}_{n,q,m}\sim \mathcal{U} (a_{n,q,m}-\delta^{a}_{n,q,m}, a_{n,q,m}+\delta^{a}_{n,q,m})$ with the half disturbation range $\delta^{a}_{n,q,m}$. Without loss of generality, we herein assume both the small-scale
fading error and the steering vector elements satisfy the uniform distribution with $\mathcal{U} (a, b)$ denoting the uniform distribution on range $[a, b]$.

We consider the error disturbation magnitudes $\delta^{\eta}_{n,q,k}={\eta}_{n,q,k}p\%$  and $\delta^{a}_{n,q,m}={a}_{n,q,m}p\%$ with the percentage $p$ given by 5 or 10. The simulation results are demonstrated in Figs. 16 and 17. The legend 'DER, 5, 5' denotes the DER performance with $5\%$ disturbation for the random fading and $5\%$ disturbation for the steering vector elements, respectively. For the proposed J-ABF-SP algorithm, the negligible DER performance degradation and the comparably large SER performance loss can be observed due to the CSI error. In addition, the SER performance deterioration would be incurred by the CSI error for the interference cancellation-based scheme (J-ABF-SP-IC) because the involved interference reconstruction relies on the CSI estimation. Overall, the performance degradation lies in an acceptable level, even with relatively large CSI errors.

\begin{figure}[!t]
   \centerline
   {\includegraphics[width=0.4\textwidth]{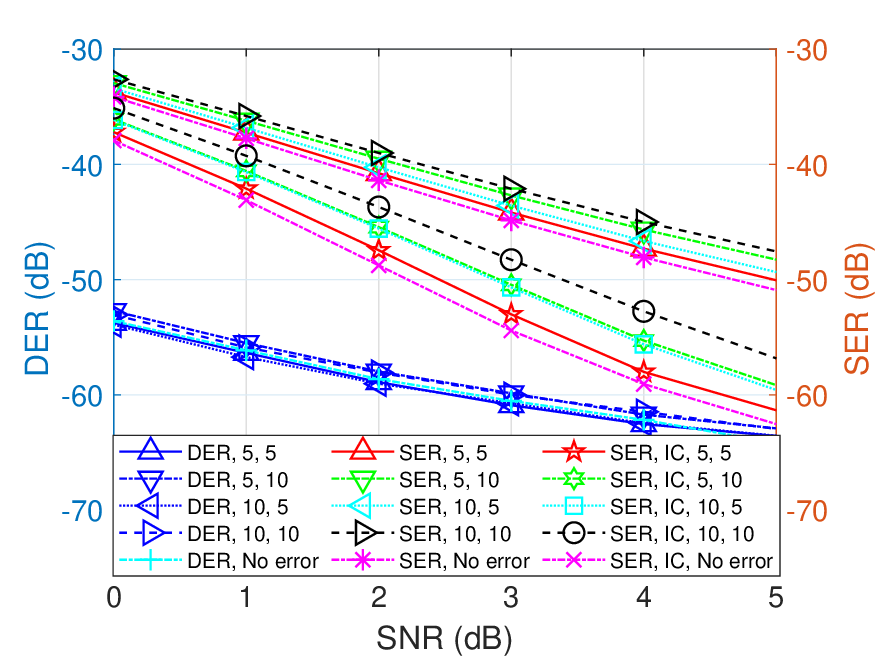}}
   \caption{The performance robustness to the CSI error under different input SNRs}\label{fig:SNR_CSIErr}
   \vspace{-0.3cm}
\end{figure}
\begin{figure}[!t]
   \centerline
   {\includegraphics[width=0.4\textwidth]{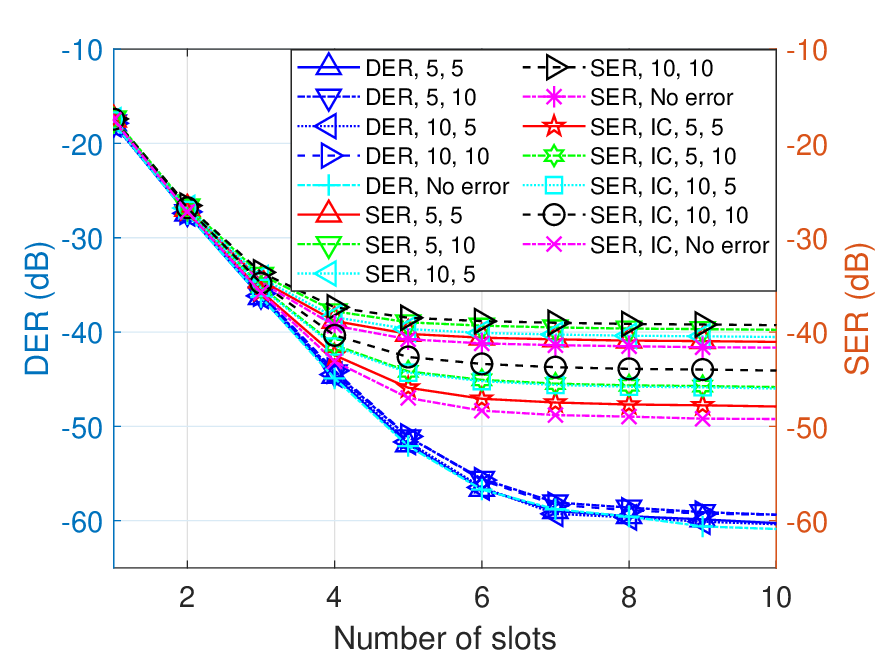}}
   \caption{The performance robustness to the CSI error under different frame length}\label{fig:Slots_CSIErr}
   \vspace{-0.3cm}
\end{figure}

\section{Conclusion and Future Work}
In this paper, we presented a general framework for the integration of the SDMA with the CS-based grant-free NOMA
for the mMTC. Two beamforming schemes were proposed for the realisation of SDMA. In particular,
we developed a joint adaptive beamforming and subspace pursuit algorithm for the user detection and data recovery, with a novel user sparsity decision method without knowing the noise level. We also devised an interference cancellation
scheme to further enhance the data recovery performance.

In the future, we will study the amalgamation of the SDMA and CS for the dynamic user sparsity-based grant-free
NOMA. To reduce the complexity, we will also study the computationally efficient CS method for the user detection and
data recovery.

\appendices
\section{Channel correlation coefficient}\label{sec:channel correlation}
The channel correlation between any two users is defined by the Pearson correlation coefficient, i.e.,
\begin{align}
   \rho_{q,p}\triangleq\dfrac{|({\bm g}_q-\bar{g}_q)^{\rm H}({\bm g}_p-\bar{g}_p)|}{\|{\bm g}_q-\bar{g}_q\|_2\|{\bm g}_p-\bar{g}_p\|_2}. \label{eq:def_corr}
\end{align}
where $\bar{g}_q$ and $\bar{g}_p$ are the average values of all the elements in vector ${\bm g}_q$ and ${\bm g}_p$, respectively. In our work, the channel gain vector is defined as the product of the channel fading and the steering vector, i.e., ${\bm g}_{n,q,k} = f_{n,q,k}{\bm a}_{n,q}$. In fact, we can approximately substitute the average values in \eqref{eq:def_corr} with zeros since the channel fading factor $f_{n,q,k}=\rho_{n,q}\eta_{n,q,k}$ follows the complex Gaussian distribution with zero mean. Therefore, the channel correlation coefficient can be given by,
\begin{align}
   \rho_{nq,lp}\triangleq{|{\bm a}_{n,q}^{\rm H}{\bm a}_{l,p}|}/{M}.\label{eq:def_corr_channel}
\end{align}
It can be viewed as the correlation between the corresponding steering vectors. Note that the channel fading factors in \eqref{eq:def_corr_channel} have been removed because they appear in both denominator and numerator.
With the steering vector defined in \eqref{eq:steering vector}, the channel correlation coefficient \eqref{eq:def_corr_channel} can be further written as,
\begin{align}
   \rho_{nq,lp}&\triangleq{|\sum_{m=0}^{M-1}e^{-j\pi m(\phi_{l,p}-\phi_{n,q})}|}/{M},\\
   &=\left|\dfrac{\sin\left(\dfrac{\pi M(\phi_{l,p}-\phi_{n,q})}{2}\right)}{M\sin\left(\dfrac{\pi(\phi_{l,p}-\phi_{n,q})}{2}\right)}\right|,\label{eq:simplified_corr_channel}
\end{align}
where $\phi_{n,q}\triangleq\dfrac{2d\sin(\theta_{n,q})}{\lambda}$. Note that \eqref{eq:simplified_corr_channel} follows from the definition of the Fej'er kernel which converges to zero quickly when its input parameter $\phi_{l,p}-\phi_{n,q}$ increases. This means that the correlation of two users' channel vectors can be measured by the normalised direction, such as $\phi_{l,p}$ and $\phi_{n,q}$. Therefore, the user clustering can be performed based on the a priori estimated channel information by the K-means method.

\section{Zadoff-Chu spreading sequences}\label{sec:Appen_Zadoff}
A ZC sequence of length $K$, consisting of $K$ complex numbers, can be denoted as ${\bm z}_q = [z_{q,0}, z_{q,1}, ..., z_{q,K-1}]^{\rm T}$. Each element of the $\beta$-root NC sequence is given by \cite{Popovic1992,Chu1972},
\begin{align}
   {s_{n,q,k}=}
   \begin{cases}
      \exp(-j \pi \beta  k (k + 1 + 2q) / K),\, \textit{K is odd},\\
      \exp(-j \pi \beta  k (k + 2q) / K),\, \textit{K is even},
   \end{cases} \label{eq:spreading sequence}
\end{align}
where $K$ is the length of the sequence, $k=0,1,\cdots,K-1$ is the index of the element in the sequence, root index $\beta$, coprime to $K$, satisfies $0 < \beta < K$, and the shift index $q$ can be any integer. In our work, we formulate the ZC spreading signature for user $u_{n,q}$ by $s_{n,q,k} = s_{q,k}$, where
$n = 1, 2, \cdots, N$ is the user cluster index and $q = 1, 2, \cdots, Q$
denotes the user index. We have the spreading signature vector $s_{n,q} = [s_{n,q,1}, s_{n,q,2}, \cdots, s_{n,q,K}]^{\rm T}$. For simplicity, we have expressed the spreading signature vector of each user by its
index $q$, while in fact, $Q$ spreading vectors can be randomly assigned to the $Q$ users according to the permutation of the user indexes.
\section{The Moore-Penrose Inverse of a Block Matrix with a Full Column Rank}\label{sec:Appen_MP_inv}
We now present a method for solving the M-P inverse of a block matrix with a full column rank. We first consider a complex-valued block matrix with a full column rank, i.e., ${\bm C}=\begin{bmatrix}{\bm A}&{\bm B}\end{bmatrix}$ where both ${\bm A}\in\mathbb{C}^{M\times n}$ and ${\bm B}\in\mathbb{C}^{M\times q}$ are with full column ranks. Define the M-P inverse of ${\bm C}$ as ${\bm C}^\dagger =\begin{bmatrix}{\bm A}^\dagger-{\bm F}\\{\bm W}^{\rm H}\end{bmatrix}$,
where ${\bm F}\in\mathbb{C}^{n\times M}$ and ${\bm W}\in\mathbb{C} ^{M\times q}$ are matrices to be determined by using the known ${\bm A}$ and ${\bm B}$.
According to ${\bm C}^\dagger{\bm C}={\bm I}$, we have
\begin{align}
   &{\bm F}{\bm A}={\bm 0},\label{eq:cond1}\\
   &({\bm A}^\dagger-{\bm F}){\bm B}={\bm 0},\label{eq:cond2}\\
   &{\bm W}^{\rm H}{\bm A}={\bm 0},\label{eq:cond3}\\
   &{\bm W}^{\rm H}{\bm B}={\bm I}.\label{eq:cond4}
\end{align}
We define ${\bm F}={\bm G}{\bm W}^{\rm H}$ with any matrix ${\bm G}\in\mathbb{C} ^{n\times q}$. In this case, \eqref{eq:cond3} leads to \eqref{eq:cond1}. Then, according to \eqref{eq:cond2} and \eqref{eq:cond4}, we have ${\bm G}={\bm A}^\dagger{\bm B}$ and thus ${\bm F}={\bm A}^\dagger{\bm B}{\bm W}^{\rm H}$. 

Subsequently, we need to solve ${\bm W}$ from \eqref{eq:cond3} and \eqref{eq:cond4}. From \eqref{eq:cond3}, we can find a matrix ${\bm U}=({\bm D}+{\bm B})-{\bm A}{\bm A}^\dagger({\bm D}+{\bm B})\in\mathbb{C}^{M\times q}$ satisfying ${\bm U}^{\rm H}{\bm A}={\bm 0}$ where ${\bm D}$ is any matrix with matching dimensions and we have used $({\bm A}{\bm A}^\dagger)^{\rm H}={\bm A}{\bm A}^\dagger$ and ${\bm A}{\bm A}^\dagger{\bm A}={\bm A}$. We define ${\bm W}={\bm U}{\bm J}$ with unknown ${\bm J}$. According to \eqref{eq:cond4}, we have,
\begin{align}
   {\bm J}^{\rm H}{\bm U}^{\rm H}{\bm B}={\bm I}&\Rightarrow {\bm J}^{\rm H}{\bm U}^{\rm H}({\bm U}-{\bm D}+{\bm A}{\bm A}^\dagger({\bm D}+{\bm B}))={\bm I}\notag\\
   &\Rightarrow {\bm J}^{\rm H}{\bm U}^{\rm H}{\bm U}-{\bm J}^{\rm H}{\bm U}^{\rm H}{\bm D}={\bm I}
\end{align}  
We can easily find ${\bm D}={\bm 0}$ and ${\bm J}=({\bm U}^{\rm H}{\bm U})^{-1}$ are the solutions. Thus, we have ${\bm W}={\bm U}({\bm U}^{\rm H}{\bm U})^{-1}$ with ${\bm U}={\bm B}-{\bm A}{\bm A}^\dagger{\bm B}$.

\section{The Monotonous Decreasing of the Residual Energy regarding the Sparsity Level}\label{sec:Appen_residual}
We now verify the {\it monotonous decreasing of the residual energy} with the sparsity level increasing up to the real one. 
With the stopping condition for beamforming update reached, the residual energy for the sparsity $s$ can be derived in light of \eqref{eq:LS-problem}, \eqref{eq:IpNC estimate}-\eqref{eq:est_Rn},
\begin{align}
   \hat{\varepsilon}_s&=\sum\nolimits_{k=1}^K\sum\nolimits_{t=1}^\mathcal{T} \hat{\bm b}_n^{\rm H}\hat{\bm i}_{n,k,t}\hat{\bm i}_{n,k,t}^{{\rm H}}\hat{\bm b}_n=K\mathcal{T}{\hat{\bm b}_n^{\rm H}\hat{\bm R}_n\hat{\bm b}_n},\label{eq:residual energy}
\end{align}
where the estimated IpNC by \eqref{eq:IpNC estimate} can be rewritten as,
\begin{align}
\hat{\bm i}_{n,k,t}={\bm i}_{n,k,t}+\tilde{\bm G}_{n,k}\tilde{\bm x}_{n,t},\label{eq:IpNC estimate simplify}
\end{align}
with the IpNC ${\bm i}_{n,k,t}$ defined in \eqref{eq:real IpNC} and $\tilde{\bm x}_{n,t}={\bm x}_{n,t}-\hat{\bm x}_{n,t}$. 

With $s<s_o$, the signal estimate $\hat{\bm x}_{n,t}$ by \eqref{eq:signal estimate normal} is inaccurate due to the UDAUs and IpNC. It consists of three parts at any $t$, i.e., 
$\hat{\bm x}_{n,t}[\varGamma_{s},1]\neq {\bm 0}$, $\hat{\bm x}_{n,t}[\tilde{\varGamma}_n\setminus \varGamma_{s},1]={\bm 0}$, and $\hat{\bm x}_{n,t}[\mathcal{Q} \setminus (\tilde{\varGamma}_n\cup \varGamma_{s}),1]={\bm 0}$. 
Then, we have the estimation error $\tilde{\bm x}_{n,t}$, i.e.,
$\tilde{\bm x}_{n,t}[\varGamma_{s},1]={\bm x}_{n,t}[\varGamma_{s},1]-\hat{\bm x}_{n,t}[\varGamma_{s},1]$, $\tilde{\bm x}_{n,t}[\tilde{\varGamma}_n\setminus \varGamma_{s},1]={\bm x}_{n,t}[\tilde{\varGamma}_n\setminus \varGamma_{s},1]$, and $\tilde{\bm x}_{n,t}[\mathcal{Q} \setminus (\tilde{\varGamma}_n\cup \varGamma_{s}),1]={\bm 0}$.
Thus, the IpNC estimate $\hat{\bm i}_{n,k,t}$ in \eqref{eq:IpNC estimate simplify} contains the residual signal component of the DAUs, the signal component of the UDAUs and the real IpNC.
The suppression on the signal component of UDAUs in $\hat{\bm i}_{n,k,t}$ is much smaller than that on the IpNC due to the beam constraint $\hat{\bm b}_n^{\rm H}\bar{\bm a}_n=1$. Thus, the residual energy $\varepsilon_s$ in \eqref{eq:residual energy} with $s<s_o$ mainly consists of the signal component of UDAUs followed by the suppressed IpNC. 

As $s$ increases, the number of the UDAUs decreases. Hence, the signal component of the UDAUs in $\hat{\bm i}_{n,k,t}$ is weakened. Meantime, the suppression for ${\bm i}_{n,k,t}$ by beamforming can be enhanced. Therefore, the residual energy $\varepsilon_s$ will gradually decrease with the given sparsity $s$ increasing up to $s_o$. 
\IEEEpeerreviewmaketitle

\small
\bibliographystyle{IEEEtran}
\bibliography{mybibfile_response}

\begin{thebibliography}{10}
\providecommand{\url}[1]{#1}
\csname url@samestyle\endcsname
\providecommand{\newblock}{\relax}
\providecommand{\bibinfo}[2]{#2}
\providecommand{\BIBentrySTDinterwordspacing}{\spaceskip=0pt\relax}
\providecommand{\BIBentryALTinterwordstretchfactor}{4}
\providecommand{\BIBentryALTinterwordspacing}{\spaceskip=\fontdimen2\font plus
\BIBentryALTinterwordstretchfactor\fontdimen3\font minus
  \fontdimen4\font\relax}
\providecommand{\BIBforeignlanguage}[2]{{%
\expandafter\ifx\csname l@#1\endcsname\relax
\typeout{** WARNING: IEEEtran.bst: No hyphenation pattern has been}%
\typeout{** loaded for the language `#1'. Using the pattern for}%
\typeout{** the default language instead.}%
\else
\language=\csname l@#1\endcsname
\fi
#2}}
\providecommand{\BIBdecl}{\relax}
\BIBdecl

\bibitem{Hoshyar2008}
R.~Hoshyar, F.~P. Wathan, and R.~Tafazolli, ``Novel low-density signature for
  synchronous {CDMA} systems over {AWGN} channel,'' \emph{IEEE Trans. Signal
  Process.}, vol.~56, no.~4, pp. 1616--1626, 2008.

\bibitem{Nikopour2013}
H.~Nikopour and H.~Baligh, ``Sparse code multiple access,'' in \emph{2013 IEEE
  24th Ann. Int. Symp. Personal, Indoor, and Mobile Radio Commun. (PIMRC)},
  2013, pp. 332--336.

\bibitem{Zhang2014}
S.~Zhang, X.~Xu, L.~Lu, Y.~Wu, G.~He, and Y.~Chen, ``Sparse code multiple
  access: An energy efficient uplink approach for {5G} wireless systems,'' in
  \emph{2014 IEEE Global Commun. Conf.}, 2014, pp. 4782--4787.

\bibitem{Islam2017}
S.~M.~R. Islam, N.~Avazov, O.~A. Dobre, and K.-s. Kwak, ``Power-domain
  non-orthogonal multiple access ({NOMA}) in {5G} systems: Potentials and
  challenges,'' \emph{IEEE Commun. Surveys $\&$ Tutorials}, vol.~19, no.~2, pp.
  721--742, 2017.

\bibitem{Kusaladharma2021}
S.~Kusaladharma, W.-P. Zhu, W.~Ajib, and G.~A.~A. Baduge, ``Achievable rate
  characterization of {NOMA}-aided cell-free massive mimo with imperfect
  successive interference cancellation,'' \emph{IEEE Trans. Commun.}, vol.~69,
  no.~5, pp. 3054--3066, 2021.

\bibitem{Shariatmadari2015}
H.~Shariatmadari, R.~Ratasuk, S.~Iraji, A.~Laya, T.~Taleb, R.~Jäntti, and
  A.~Ghosh, ``Machine-type communications: current status and future
  perspectives toward {5G} systems,'' \emph{IEEE Commun. Mag.}, vol.~53, no.~9,
  pp. 10--17, 2015.

\bibitem{3GPP_GF}
Lenovo, ``Uplink grant-free access for {5G} {mMTC},'' \emph{3GPP document
  R1-1609398, TSG-RAN WG1 Meeting $\#86b$}, October 2016.

\bibitem{Dawy2017}
Z.~Dawy, W.~Saad, A.~Ghosh, J.~G. Andrews, and E.~Yaacoub, ``Toward massive
  machine type cellular communications,'' \emph{IEEE Wireless Commun.},
  vol.~24, no.~1, pp. 120--128, 2017.

\bibitem{Liu2018MSP}
L.~Liu, E.~G. Larsson, W.~Yu, P.~Popovski, C.~Stefanovic, and E.~de~Carvalho,
  ``Sparse signal processing for grant-free massive connectivity: A future
  paradigm for random access protocols in the internet of things,'' \emph{IEEE
  Signal Process. Mag.}, vol.~35, no.~5, pp. 88--99, 2018.

\bibitem{Shahab2020}
M.~B. Shahab, R.~Abbas, M.~Shirvanimoghaddam, and S.~J. Johnson, ``Grant-free
  non-orthogonal multiple access for {IoT}: A survey,'' \emph{IEEE Commun.
  Surveys and Tutorials}, vol.~22, no.~3, pp. 1805--1838, 2020.

\bibitem{Qiao2022}
L.~Qiao, J.~Zhang, Z.~Gao, D.~Zheng, M.~J. Hossain, Y.~Gao, D.~W.~K. Ng, and
  M.~Di~Renzo, ``Joint activity and blind information detection for
  {UAV}-assisted massive {IoT} access,'' \emph{IEEE J. Sel. Areas Commun.},
  vol.~40, no.~5, pp. 1489--1508, 2022.

\bibitem{Senel2018}
K.~Senel and E.~G. Larsson, ``Grant-free massive {MTC}-enabled massive {MIMO}:
  A compressive sensing approach,'' \emph{IEEE Trans. Commun.}, vol.~66,
  no.~12, pp. 6164--6175, 2018.

\bibitem{Liu2018}
L.~Liu and W.~Yu, ``Massive connectivity with massive mimo—part i: Device
  activity detection and channel estimation,'' \emph{IEEE Trans. Signal
  Process.}, vol.~66, no.~11, pp. 2933--2946, 2018.

\bibitem{Ahn2019}
J.~Ahn, B.~Shim, and K.~B. Lee, ``{EP}-based joint active user detection and
  channel estimation for massive machine-type communications,'' \emph{IEEE
  Trans. Commun.}, vol.~67, no.~7, pp. 5178--5189, 2019.

\bibitem{Ke2020}
M.~Ke, Z.~Gao, Y.~Wu, X.~Gao, and R.~Schober, ``Compressive sensing-based
  adaptive active user detection and channel estimation: Massive access meets
  massive {MIMO},'' \emph{IEEE Trans. Signal Process.}, vol.~68, pp. 764--779,
  2020.

\bibitem{Wei2017AMP}
C.~Wei, H.~Liu, Z.~Zhang, J.~Dang, and L.~Wu, ``Approximate message
  passing-based joint user activity and data detection for {NOMA},'' \emph{IEEE
  Commun. Lett.}, vol.~21, no.~3, pp. 640--643, 2017.

\bibitem{Du2018}
Y.~Du, C.~Cheng, B.~Dong, Z.~Chen, X.~Wang, J.~Fang, and S.~Li,
  ``Block-sparsity-based multiuser detection for uplink grant-free {NOMA},''
  \emph{IEEE Trans. Wireless Commun.}, vol.~17, no.~12, pp. 7894--7909, 2018.

\bibitem{Gao2022}
P.~Gao, Z.~Liu, P.~Xiao, C.~H. Foh, and J.~Zhang, ``Low-complexity block
  coordinate descend based multiuser detection for uplink grant-free {NOMA},''
  \emph{IEEE Trans. Veh. Technology}, pp. 1--1, 2022.

\bibitem{Mei2022}
Y.~Mei, Z.~Gao, Y.~Wu, W.~Chen, J.~Zhang, D.~W.~K. Ng, and M.~Di~Renzo,
  ``Compressive sensing-based joint activity and data detection for grant-free
  massive {IoT} access,'' \emph{IEEE Trans. Wireless Commun.}, vol.~21, no.~3,
  pp. 1851--1869, 2022.

\bibitem{Tropp2007}
J.~A. Tropp and A.~C. Gilbert, ``Signal recovery from random measurements via
  orthogonal matching pursuit,'' \emph{IEEE Trans. Inf. Theory}, vol.~53,
  no.~12, pp. 4655--4666, 2007.

\bibitem{Guo2008}
D.~Guo and C.-c. Wang, ``Multiuser detection of sparsely spread {CDMA},''
  \emph{IEEE J. Sel. Areas Commun.}, vol.~26, no.~3, pp. 421--431, 2008.

\bibitem{Needell2009}
D.~Needell and J.~Tropp, ``{CoSaMP}: iterative signal recovery from incomplete
  and inaccurate samples,'' \emph{Communications of the ACM}, vol.~53, no.~12,
  pp. 93--100, 2010.

\bibitem{Dai2009}
W.~Dai and O.~Milenkovic, ``Subspace pursuit for compressive sensing signal
  reconstruction,'' \emph{IEEE Trans. Inf. Theory}, vol.~55, no.~5, pp.
  2230--2249, 2009.

\bibitem{donoho2009message}
D.~L. Donoho, A.~Maleki, and A.~Montanari, ``Message-passing algorithms for
  compressed sensing,'' \emph{Proceedings of the National Academy of Sciences},
  vol. 106, no.~45, pp. 18\,914--18\,919, 2009.

\bibitem{Luo2022}
Q.~Luo, Z.~Liu, G.~Chen, Y.~Ma, and P.~Xiao, ``A novel multitask learning
  empowered codebook design for downlink {SCMA} networks,'' \emph{IEEE Wireless
  Commun. Lett.}, vol.~11, no.~6, pp. 1268--1272, 2022.

\bibitem{Luo2022Non_coherent}
Q.~Luo, H.~Wen, G.~Chen, Z.~Liu, P.~Xiao, Y.~Ma, and A.~Maaref, ``A novel
  non-coherent {SCMA} with massive {MIMO},'' \emph{IEEE Wireless Commun.
  Lett.}, pp. 1--1, 2022.

\bibitem{Wang2016}
B.~Wang, L.~Dai, Y.~Zhang, T.~Mir, and J.~Li, ``Dynamic compressive
  sensing-based multi-user detection for uplink grant-free {NOMA},'' \emph{IEEE
  Commun. Lett.}, vol.~20, no.~11, pp. 2320--2323, 2016.

\bibitem{Du2017}
Y.~Du, B.~Dong, Z.~Chen, X.~Wang, Z.~Liu, P.~Gao, and S.~Li, ``Efficient
  multi-user detection for uplink grant-free {NOMA}: Prior-information aided
  adaptive compressive sensing perspective,'' \emph{IEEE J. Sel. Areas
  Commun.}, vol.~35, no.~12, pp. 2812--2828, 2017.

\bibitem{Li2022}
T.~Li, J.~Zhang, Z.~Yang, Z.~L. Yu, Z.~Gu, and Y.~Li, ``Dynamic user activity
  and data detection for grant-free {NOMA} via weighted $l_{2,1}$
  minimization,'' \emph{IEEE Trans. Wireless Commun.}, vol.~21, no.~3, pp.
  1638--1651, 2022.

\bibitem{Wu2022}
L.~Wu, Z.~Wang, P.~Sun, and Y.~Yang, ``Temporal correlation enhanced sparse
  activity detection in {MIMO} enabled grant-free noma,'' \emph{IEEE Trans.
  Veh. Technology}, vol.~71, no.~3, pp. 2887--2899, 2022.

\bibitem{Wu2021}
L.~Wu, P.~Sun, Z.~Wang, and Y.~Yang, ``Joint user activity identification and
  channel estimation for grant-free {NOMA}: A spatial–temporal
  structure-enhanced approach,'' \emph{IEEE Internet of Things J.}, vol.~8,
  no.~15, pp. 12\,339--12\,349, 2021.

\bibitem{Ma2019}
X.~Ma, J.~Kim, D.~Yuan, and H.~Liu, ``Two-level sparse structure-based
  compressive sensing detector for uplink spatial modulation with massive
  connectivity,'' \emph{IEEE Commun. Lett.}, vol.~23, no.~9, pp. 1594--1597,
  2019.

\bibitem{Qiao2020}
L.~Qiao, J.~Zhang, Z.~Gao, S.~Chen, and L.~Hanzo, ``Compressive sensing based
  massive access for iot relying on media modulation aided machine type
  communications,'' \emph{IEEE Trans. Veh. Technology}, vol.~69, no.~9, pp.
  10\,391--10\,396, 2020.

\bibitem{Qiao2022MM}
L.~Qiao, J.~Zhang, Z.~Gao, D.~W.~K. Ng, M.~D. Renzo, and M.-S. Alouini,
  ``Massive access in media modulation based massive machine-type
  communications,'' \emph{IEEE Trans. Wireless Commun.}, vol.~21, no.~1, pp.
  339--356, 2022.

\bibitem{Ma2020}
X.~Ma, S.~Guo, and D.~Yuan, ``Improved compressed sensing-based joint user and
  symbol detection for media-based modulation-enabled massive machine- type
  communications,'' \emph{IEEE Access}, vol.~8, pp. 70\,058--70\,070, 2020.

\bibitem{Xu2017}
Y.~Xu, C.~Shen, Z.~Ding, X.~Sun, S.~Yan, G.~Zhu, and Z.~Zhong, ``Joint
  beamforming and power-splitting control in downlink cooperative {SWIPT NOMA}
  systems,'' \emph{IEEE Trans. Signal Process.}, vol.~65, no.~18, pp.
  4874--4886, 2017.

\bibitem{Cui2018}
J.~Cui, Z.~Ding, P.~Fan, and N.~Al-Dhahir, ``Unsupervised machine
  learning-based user clustering in millimeter-wave-{NOMA} systems,''
  \emph{IEEE Trans. Wireless Commun.}, vol.~17, no.~11, pp. 7425--7440, 2018.

\bibitem{Zhu2019}
L.~Zhu, J.~Zhang, Z.~Xiao, X.~Cao, D.~O. Wu, and X.-G. Xia, ``Joint {Tx-Rx}
  beamforming and power allocation for {5G} millimeter-wave non-orthogonal
  multiple access networks,'' \emph{IEEE Trans. Commun.}, vol.~67, no.~7, pp.
  5114--5125, 2019.

\bibitem{Senel2019}
K.~Senel, H.~V. Cheng, E.~Björnson, and E.~G. Larsson, ``What role can {NOMA}
  play in massive {MIMO}?'' \emph{IEEE J. Sel. Topics Signal Process.},
  vol.~13, no.~3, pp. 597--611, 2019.

\bibitem{Zhang2020}
H.~Zhang, H.~Zhang, W.~Liu, K.~Long, J.~Dong, and V.~C.~M. Leung, ``Energy
  efficient user clustering, hybrid precoding and power optimization in
  terahertz {MIMO-NOMA} systems,'' \emph{IEEE J. Select. Areas Commun.},
  vol.~38, no.~9, pp. 2074--2085, 2020.

\bibitem{Xia2022}
G.~Xia, Y.~Zhang, L.~Ge, and H.~Zhou, ``Deep reinforcement learning based
  dynamic power allocation for uplink device-to-device enabled cell-free
  network,'' in \emph{2022 IEEE Int. Symp. Broadband Multimedia Syst. and
  Broadcast. (BMSB)}, 2022, pp. 01--06.

\bibitem{Le2021}
Q.~N. Le, V.-D. Nguyen, O.~A. Dobre, N.-P. Nguyen, R.~Zhao, and S.~Chatzinotas,
  ``Learning-assisted user clustering in cell-free massive {MIMO-NOMA}
  networks,'' \emph{IEEE Trans. Veh. Technol.}, vol.~70, no.~12, pp.
  12\,872--12\,887, 2021.

\bibitem{Popovic1992}
B.~Popovic, ``Generalized chirp-like polyphase sequences with optimum
  correlation properties,'' \emph{IEEE Trans. Inform. Theory}, vol.~38, no.~4,
  pp. 1406--1409, 1992.

\bibitem{Van1988}
B.~Van~Veen and K.~Buckley, ``Beamforming: a versatile approach to spatial
  filtering,'' \emph{IEEE ASSP Mag.}, vol.~5, no.~2, pp. 4--24, 1988.

\bibitem{3GPPPCM}
``Evolved universal terrestrial radio access ({E-UTRA}); physical channels and
  modulation (release 12),'' \emph{3GPP document TS-36.211}, January 2016.

\bibitem{Zhang2019JCN}
Y.~Zhang, H.~Cao, M.~Zhou, and L.~Yang, ``Cell-free massive {MIMO}: Zero
  forcing and conjugate beamforming receivers,'' \emph{J. Commun. Networks},
  vol.~21, no.~6, pp. 529--538, 2019.

\bibitem{Chu1972}
D.~Chu, ``Polyphase codes with good periodic correlation properties
  (corresp.),'' \emph{IEEE Trans. Inform. Theory}, vol.~18, no.~4, pp.
  531--532, 1972.

\end{thebibliography}

\end{document}